\documentclass[preprint,prd,aps,superscriptaddress,11pt]{revtex4-1}

\usepackage{graphicx}
\usepackage{amsmath}
\usepackage{amssymb}
\usepackage{dcolumn}
\usepackage{bm}
\usepackage{slashed}
\usepackage{color}
\usepackage{changepage}
\usepackage{multirow}
\usepackage{hyperref}

\newcommand{\be}{\begin{equation}}
\newcommand{\ee}{\end{equation}}
\newcommand{\ba}{\begin{eqnarray}}
\newcommand{\ea}{\end{eqnarray}}
\newcommand{\no}{\nonumber \\}
\newcommand{\gsim}{\mathrel{\hbox{\rlap{\lower.55ex \hbox {$\sim$}}
                   \kern-.3em \raise.4ex \hbox{$>$}}}}
\newcommand{\lsim}{\mathrel{\hbox{\rlap{\lower.55ex \hbox {$\sim$}}
                   \kern-.3em \raise.4ex \hbox{$<$}}}}

\def\roughly#1{\mathrel{\raise.3ex\hbox{$#1$\kern-.75em%
\lower1ex\hbox{$\sim$}}}}
\def\lsim{\roughly<}
\def\gsim{\roughly>}

\def\({\left(}
\def\){\right)}
\def\[{\left[}
\def\]{\right]}
\def\<{\langle}
\def\>{\rangle}

\def\tu{{\tilde u}}
\def\tb{{\tilde b}}

\def\cJ{{\cal J}}

\def\cT{{\cal T}}
\def\cL{{\cal L}}

\def\cD{{\cal D}}
\def\cV{{\cal V}}
\def\cA{{\cal A}}
\def\cB{{\cal B}}

\def\cG{{\cal G}}
\def\pb{{\bar p}}
\def\qb{{\bar q}}
\def\ab{{\bar a}}
\def\fb{{\bar f}}

\def\Lb{{\bar \L}}
\def\hp{{\hat p}}
\def\qh{{\hat q}}

\def\gh{{\hat g}}

\def\l{{\lambda}}
\def\L{{\Lambda}}
\def\d{{\delta}}
\def\D{{\Delta}}
\def\o{{\omega}}

\def\e{{\epsilon}}

\def\a{{\alpha}}
\def\b{{\beta}}
\def\c{{\chi}}

\def\g{{\gamma}}
\def\G{{\Gamma}}
\def\p{{\pi}}

\def\m{{\mu}}
\def\n{{\nu}}
\def\r{{\rho}}
\def\s{{\sigma}}

\def\th{{\theta}}

\def\hb{{\hbar}}

\newcommand{\pd}{\partial}
\newcommand{\pdb}{\bar \pd}
\newcommand{\pr}{\parallel}
\newcommand{\pp}{\perp}

\date{\today}

\begin{document}

\title{\bf Two-point functions from chiral kinetic theory in magnetized plasma}

\author{Lixin Yang}
\email{yanglx5@mail2.sysu.edu.cn}
\affiliation{School of Physics and Astronomy, Sun Yat-Sen University, Zhuhai 519082, China}

\begin{abstract}

We study the two-point functions from chiral kinetic theory which characterize the response to perturbative vector and axial gauge fields in magnetized chiral plasma. In the lowest Landau level approximation, the solution of chiral kinetic equations gives density waves of electric and axial charges, which contain chiral magnetic wave implied by the axial anomaly and magnetic field. We then obtain the constitutive relations for covariant currents and stress tensor that involving the density waves. By considering the difference between consistent and covariant anomalies explicitly, the correlators of consistent currents and stress tensor satisfy derivative symmetry, and therefore allow an effective action for the perturbative gauge fields as the generating functional of the correlators. We also verify the derivative symmetry of the correlators agrees with the Onsager relations. 

\end{abstract}

\maketitle


\newpage

\section{Introduction}

It has been revealed that axial anomaly signifies a type of gapless collective excitation arising from the coupling of the the electric and chiral charge density waves, which is known as the chiral magnetic wave (CMW) \cite{Kharzeev:2010gd}. These density waves are closely related to the anomalous transports of chiral magnetic effect (CME) \cite{Vilenkin:1980fu,Kharzeev:2004ey,Kharzeev:2007tn,Fukushima:2008xe,Son:2009tf,Neiman:2010zi} and chiral separation effect (CSE) \cite{Metlitski:2005pr,Son:2004tq}. For the quark-gluon plasma with a strong magnetic field, it has been demonstrated in a dimensionally reduced (1+1)D theory \cite{Kharzeev:2010gd} that the CMW indeed stems from the connections $J_0=J_1^5$, $J_0^5=J_1$ between the vector $J$ and axial $J^5$ currents of CME and CSE. Besides the transports along magnetic field, axial anomaly also implies anomalous currents from chiral vortical effects (CVE) \cite{Vilenkin:1980zv,Erdmenger:2008rm,Banerjee:2008th,Son:2009tf,Neiman:2010zi,Landsteiner:2011cp} as responses to vorticity. 

The chiral anomalous effects have been investigated in the framework of chiral kinetic theory (CKT) \cite{Son:2012wh,Son:2012zy,Stephanov:2012ki,Gao:2012ix,Pu:2010as,Chen:2012ca,Hidaka:2016yjf,Manuel:2013zaa,Manuel:2014dza,Wu:2016dam,Mueller:2017arw,Mueller:2017lzw,Huang:2018wdl,Gao:2018wmr,Carignano:2018gqt,Lin:2019ytz,Carignano:2019zsh,Liu:2018xip,Weickgenannt:2019dks,Gao:2019znl,Hattori:2019ahi,Wang:2019moi,Yang:2020hri,Liu:2020flb,Hayata:2020sqz,chen2021equaltime}, which can be viewed as an expansion in weak electromagnetic field and spacetime gradient. Furthermore, CKT with Landau level basis \cite{Lin:2019fqo,Gao:2020ksg,Hattori:2016lqx,Sheng:2017lfu,Fukushima:2019ugr} has been derived in a different expansion scheme to study plasma magnetized by non-perturbative magnetic field. Particularly, in a strong magnetic field, the CKT in lowest Landau level approximation (LLLA) and collisionless limit has been applied to study photon self-energy \cite{Gao:2020ksg} and magneto-vortical effect \cite{Lin:2021sjw}. Recently, CKT in weak field expansion has been generalized to incorporate axial gauge field to derive correlation functions of axial currents and stress tensor \cite{Chen:2021azy}. It would be interesting to study the transport phenomena in magnetized plasma as responses to perturbative vector and axial gauge fields with CKT. 

In linear regime, the transport coefficients are related to the retarded two-point correlation functions via the Kubo formulas. 
The two-point function is important in its own right, e.g. the photon self-energy as a fundamental quantity characterizing the vacuum polarization by electromagnetic field has been intensively studied for magnetized plasma in field theory \cite{Danielsson:1995rh,PhysRevD.83.111501,Chao:2014wla,Fukushima:2015wck,Chao:2016ysx,Hattori:2017xoo,Singh:2020fsj,Wang:2021eud}. On the other hand, as a contrast to the full non-perturbative effective Lagrangian of vacuum for constant electromagnetic field established by Heisenberg and Euler \cite{Heisenberg:1936nmg} in the study of vacuum polarization, an effective action for the perturbative gauge fields in magnetized medium can also be constructed as the generating functional of the correlators. 

In this paper, we introduce perturbative vector and axial gauge potentials to CKT in a strong background magnetic field to study the correlators among currents $J^\m ,\,J_5^\m$ and energy-momentum tensor $T^{\m\n}$. The CKT results, which express $J^\m ,\,J_5^\m$ and $T^{\m\n}$ in terms of the thermodynamic variables, give the constitutive relations that involve CMW mode as a consequence of axial anomaly and the presence of magnetic field. It is known that there is distinction between consistent and covariant anomalies \cite{Bardeen:1984pm}, see also \cite{2016AcPPB..47.2617L} for a review, which is more explicit in the presence of axial gauge field. We find that CKT gives covariant currents, while the consistent currents are obtained by adding the Chern-Simons currents. The correlators of the consistent currents and stress tensor satisfy the derivative symmetry upon the interchange of the functional derivatives. It turns out that the derivative symmetry of correlators also fit the Onsager relations \cite{Hernandez:2017mch}. 

The paper is organized as follows:  In Section \eqref{sec_anomaly}, we discuss the property of current-current correlators in consistent and covariant anomalies with perturbative gauge potentials and strong background magnetic field. The density waves of vector and axial charges including CMW is obtained in this condition. In Section \eqref{sec_ckt}, we give a brief introduction of the covariant CKT with the same external field configuration. We then solve the chiral kinetic equations (CKE) to derive the constitutive relations of currents and stress tensor in Section \eqref{sec_solution}, where we reproduce the density waves in the currents and verify the property of the resulting correlators, whose generating functional is derived as the effective action for gauge fields.

Throughout this paper, we set $\hb=1$ and $c=1$. We take positive charge $q_f=e$ for chiral fermions and absorb electric charge $e$ into the gauge potentials. We use the notations $(x^\m)=(x_0,{\bf x}),\,(p^\m)=(p_0,{\bf p})$ for four-vectors and adopt mostly minus signature. We also define $X^{[\m}Y^{\n]}\equiv X^\m Y^\n-X^\n Y^\m$ and $X^{\{\m}Y^{\n\}}\equiv X^\m Y^\n+X^\n Y^\m$ to be concise.

\section{Correlators from effective action in axial anomaly}\label{sec_anomaly}

In the presence of axial gauge field, the difference between consistent current $\cJ^\m$ and covariant current $J^\m$ is nontrivial \cite{2016AcPPB..47.2617L}.
With the explicitly conserved vector current, the consistent anomaly writes 
\begin{align}\label{ward_kd_cs}
  \pd_\m \cJ^\m=&0\,,\no
  \pd_\m \cJ_5^\m=&-\frac{1}{48\p^2}\e^{\m\n\r\l}\(3F_{\m\n}F_{\r\l}+F_{\m\n}^5 F_{\r\l}^5\)\,.
\end{align}
For the  gauge invariant currents, one has covariant anomaly,
\begin{align}\label{ward_kd_cv}
  \pd_\m J^\m=&-\frac{1}{8\p^2}\e^{\m\n\r\l}F_{\m\n} F_{\r\l}^5\,,\no
  \pd_\m J_5^\m=&-\frac{1}{16\p^2}\e^{\m\n\r\l}\(F_{\m\n}F_{\r\l}+F_{\m\n}^5 F_{\r\l}^5\)\,.
\end{align}
The currents are related by the Chern-Simons currents as follows,
\begin{align}\label{cs_cv}
  &\cJ^\m-\frac{1}{4\p^2}\e^{\m\n\r\l}A_\n^5 F_{\r\l}=J^\m\,,\no
  &\cJ_5^\m-\frac{1}{12\p^2}\e^{\m\n\r\l}A_\n^5 F_{\r\l}^5=J_5^\m\,.
\end{align}
We study the magnetized chiral plasma with a constant fluid velocity $u^\m$ in the strong magnetic field $B^\m$. To this end, we take vanishing axial gauge field $A_5^\m=0$ and $O(1)$ magnetic field $F_{\m\n}=\pd_{[\m} A_{\n]}=\e_{\m\n\r\s}u^\r B^\s$ as the constant backgrounds. 
In the perturbations, we turn on vector gauge field $a^\l$ and axial gauge field $a_5^\l$ symmetrically at $O(a)$ which give $f_{\m\n}=\pd_{[\m} a_{\n]}=E_{[\m} u_{\n]}+\e_{\m\n\r\s}u^\r \cB^\s$ and $f_{\m\n}^5=\pd_{[\m} a_{\n]}^5=E_{[\m}^5 u_{\n]}+\e_{\m\n\r\s}u^\r \cB_5^\s$ at $O(\pd a)$. This is also the perturbation expansion scheme of the CKT in the next section. 

To $O(\pd a)$, the Ward identities in \eqref{ward_kd_cs} and \eqref{ward_kd_cv} are reduced to
\begin{align}\label{ward_kd_1}
  \pd_\m \cJ^\m=&\, \pd_\m J^\m+\frac{1}{8\p^2}\e^{\m\n\r\l}F_{\m\n} f_{\r\l}^5=0\,,\no
  \pd_\m \cJ_5^\m=&\, \pd_\m J_5^\m=-\frac{1}{8\p^2}\e^{\m\n\r\l}F_{\m\n} f_{\r\l}\,,
\end{align}
and the currents are
\begin{align}\label{cs_cv_1}
  \cJ^\m=&J^\m+\frac{1}{4\p^2}\e^{\m\n\r\l}a_\n^5 F_{\r\l}\,,\no
  \cJ_5^\m=&J_5^\m\,.
\end{align}
In the regime of strong magnetic field, one can effectively perform a dimensional reduction from (3+1)D to (1+1)D, where the orthonormal vectors $u^\m$ and $b^\m=B^\m/B$ give the directions of temporal and spatial axes severally. In this effective (1+1)D theory, the metric writes $\gh^{\m\n}=u^\m u^\n-b^\m b^\n$, while the  anti-symmetric tensor is  $\e^{\m\n}=u^{[\m}b^{\n]}$. Then, the Ward identities \eqref{ward_kd_1} to $O(\pd a)$ can be further written as 
\begin{align}\label{ward_kd_2}
  \pd_\m \cJ^\m=&\, \pd_\m J^\m-\frac{B}{4\p^2}\e^{\r\l}f_{\r\l}^5=0\,,\no
  \pd_\m \cJ_5^\m=&\, \pd_\m J_5^\m=\frac{B}{4\p^2}\e^{\r\l}f_{\r\l}\,,
\end{align}
which can be understood as the genuine (1+1)D anomalies $\frac{1}{2\p}\e^{\r\l}f_{\r\l}^5$ and $\frac{1}{2\p}\e^{\r\l}f_{\r\l}$ multiplied by the Landau-level degeneracy $\frac{B}{2\p}$. Likewise, the currents in \eqref{cs_cv_1} write
\begin{align}\label{cs_cv_2}
  \cJ^\m=&J^\m-\frac{B}{2\p^2}\e^{\m\n}a_\n^5\,,\no
  \cJ_5^\m=&J_5^\m\,.
\end{align}
Meanwhile, the dimensional reducd vector and axial currents are connected by the leading order constitutive relation,
\begin{align}\label{j5_j}
  J_5^\m=-\e^{\m\n}J_\n\,.
\end{align}
Note we have used the covariant currents in the above gauge invariant identity which will be confirmed by the CKT solution. While the consistent currents are not gauge invariant in general due to the additional gauge dependent Chern-Simons currents, a nice characteristic thereof is that they can be interpreted as the functional derivatives of effective action. The consequential current-current correlators are constrained by the derivative symmetry since the order of variations does not matter. Concretely, we have
\begin{align}\label{cor_cons}
  \cG_{VV}^{\m\n}\equiv\frac{\d\cJ^\m}{\d a_\n}=\frac{\d^2 \cL}{\d a_\n\, \d a_\m}\,,\qquad
  \cG_{AA}^{\m\n}\equiv\frac{\d\cJ_5^\m}{\d a_\n^5}=\frac{\d^2 \cL}{\d a_\n^5\, \d a_\m^5}\,,\no
  \cG_{VA}^{\m\n}\equiv\frac{\d\cJ^\m}{\d a_\n^5}=\frac{\d^2 \cL}{\d a_\n^5\, \d a_\m}\,,\qquad
  \cG_{AV}^{\m\n}\equiv\frac{\d\cJ_5^\m}{\d a_\n}=\frac{\d^2 \cL}{\d a_\n\, \d a_\m^5}\,,
\end{align}
where we have defined the correlators with effective Lagrangian density.  In momentum space, the derivative symmetry relations write
\begin{align}\label{bose_sym}
  \cG_{VV}^{\m\n}(q)=\cG_{VV}^{\n\m}(-q)\,,\qquad
  \cG_{AA}^{\m\n}(q)=\cG_{AA}^{\n\m}(-q)\,,\qquad
  \cG_{VA}^{\m\n}(q)=\cG_{AV}^{\n\m}(-q)\,,
\end{align}
which hold beyond (1+1)D.
Though the covariant currents can not be thought of as functional derivatives of effective action, we can still define the correlators in form as,
\begin{align}\label{cor_cov}
  G_{VV}^{\m\n}\equiv \frac{\d J^\m}{\d a_\n}\,,\qquad
  G_{AA}^{\m\n}\equiv \frac{\d J_5^\m}{\d a_\n^5}\,,\no
  G_{VA}^{\m\n}\equiv \frac{\d J^\m}{\d a_\n^5}\,,\qquad
  G_{AV}^{\m\n}\equiv \frac{\d J_5^\m}{\d a_\n}\,.
\end{align}
By noting the connection \eqref{cs_cv_2} to $O(\pd a)$, $\cG$ and $G$ are related as
\begin{align}\label{cor_cons_cov}
  \cG_{VV}^{\m\n}=&G_{VV}^{\m\n}\,,\qquad
  \cG_{VA}^{\m\n}=G_{VA}^{\m\n}-\frac{B}{2\p^2}u^{[\m}b^{\n]}\,,\no
  \cG_{AA}^{\m\n}=&G_{AA}^{\m\n}\,,\qquad
  \cG_{AV}^{\m\n}=G_{AV}^{\m\n}\,.
\end{align}

Before closing this section, we compute the response to the perturbative $a^\m$ and $a_5^\m$ in the effective (1+1)D theory. In such condition, there are only temporal and longitudinal components in the currents,
\begin{align}
  J^\m=n\, u^\m + n_5\, b^\m\,,\qquad
  J_5^\m=n_5\, u^\m + n\, b^\m\,.
\end{align}
Using constitutive relation \eqref{j5_j} together with the Fourier transformation of the anomaly equations \eqref{ward_kd_2}, one easily finds
\begin{align}\label{cmw}
  n=\frac{\D\m }{2\p^2}B\,,\qquad
  n_5=\frac{\D\m_5 }{2\p^2}B\,,
\end{align}
where we have defined
\begin{align}\label{che_cmw}
  \D\m\equiv&\frac{\qb_3\(\qb_3\ab_0-\qb_0\ab_3\)+\qb_0\(\qb_3\ab_0^5-\qb_0\ab_3^5\)}{\qb_0^2-\qb_3^2}\,,\no
  \D\m_5\equiv&\frac{\qb_0\(\qb_3\ab_0-\qb_0\ab_3\)+\qb_3\(\qb_3\ab_0^5-\qb_0\ab_3^5\)}{\qb_0^2-\qb_3^2}\,.
\end{align}
Here $\ab$ and $\qb$ are the gauge potentials and momentum in the local rest frame of the fluid, which can be obtained by the Lorentz tansformation defined in Appendix \eqref{app_momenta} . We interpret $\D\m$ and $\D\m_5$ as the change of chemical potential $\m$ and axial chemical potential $\m_5$ characterizing the redistribution of the chiral plasma induced by the gauge fields $a^\m$ and $a_5^\m$. Correspondingly, $n_5$ and $n$ give the Fourier transformed densities of electric and axial charges, which are related to the anomalous transports of CME and CSE. In coordinate space, $\D\m$ and $\D\m_5$ write
\begin{align}\label{che_cmw}
  \D\m=&\frac{1}{2}\cD(x-y)\[b\cdot\pd\(\e^{\m\n}f_{\m\n}\)+u\cdot\pd\(\e^{\m\n}f_{\m\n}^5\)\]\,,\no
  \D\m_5=&\frac{1}{2}\cD(x-y)\[u\cdot\pd\(\e^{\m\n}f_{\m\n}\)+b\cdot\pd\(\e^{\m\n}f_{\m\n}^5\)\]\,,
\end{align}
with $\hat \pd^\m\equiv\gh^{\m\n}\pd_\n$ and the Green's function $\cD(x-y)$ being the inverse of the operator $\hat\pd_\m\hat \pd^\m$,
\begin{align}
  \hat\pd_\m\hat \pd^\m\cD(x-y)=\int d^4y\, \d^{(4)}(x-y)\,.
\end{align}

The non-local factor $\frac{1}{\qb_0^2-\qb_3^2}$ in $n$ and $n_5$ indicates there are density waves of electric and axial charges propagate with frequency $\qb_0$ and momentum $\qb_3$ along the direction of the magnetic field. 
The poles $\qb_0 \pm \qb_3 = 0$  give the dispersion relation of the gapless collective excitation in the chiral plasma, which describes the CMW propagating with the speed of light in the limit of strong magnetic field. The poles are independent of the limit $a^\m\to0$ and $a_5^\m\to0$, which means the existence of CMW does not rely on the perturbation of the gauge potentials.

\section{Chiral kinetic theory with vector/axial gauge fields}\label{sec_ckt}

The Lorentz covariant CKE with Landau level states have been given by \cite{Lin:2021sjw} in the collisonless limit,
\begin{align}
  p_{\m} j_s^{\m}&=0\,,\label{eom1}\\
  \D_{\m} j_s^{\m}&=0\,,\label{eom2}\\
  p^{[\m} j_s^{\n]}&=-\frac{s}{2} \e^{\m\n\r\s} \D_{\r} j_{\s}^s\,,\label{eom3}
\end{align}
where $\D_\m=\pd_\m-\frac{\pd}{\pd p_\n}\(F_{\m\n}+f_{\m\n}^s\)$ with helicity $s=\pm 1$ for right/left-handed chiral fermions and $f_{\m\n}^s=f_{\m\n}+sf_{\m\n}^5$, which means the perturbative left-handed and right-handed gauge fields, or equivalently, vector and axial gauge fields are taken on the equal footing. 
We note that in \cite{Gao:2020ksg}, it has been verified that the transport equation agrees with the anomalous Ward identity. Repeating the same procedure, one readily reproduces the covariant anomaly \eqref{ward_kd_cv} from \eqref{eom2}. This implies that CKT gives covariant currents, which will be checked in the final solutions.

The CKE \eqref{eom1} to \eqref{eom3} are valid to $O(\pd a)$ expansion. Therefore we seek solution of $j_s^\m$ to the first order in perturbation $a$ and gradient $\pd a$,
\begin{align}\label{expd}
  j_s^\m=j_{(0)s}^\m+j_{a\,s}^\m+\sum_{\cA s}j_{\cA s}^\m\,,
\end{align}
where $j_{(0)s}^\m$ is the $O(1)$ background solution while $j_{a\,s}^\m$ is solution for the $O(a)$ perturbations $a_s^\m$. The last term $\sum j_{\cA s}^\m$ is the solution for the $O(\pd a)$ perturbations, where $\cA_s=E_\pr^s,E_\pp^s,\cB_\pr^s,\cB_\pp^s$ denote parallel/perpendicular electric/magnetic field respectively as is detailed in Appendix \eqref{app_EMfield}. 

The $O(1)$ background solution for $j_{(0)s}^\m$ in LLL has been given as
\begin{align}\label{cov_sol}
  j_{(0)s}^{\m}=&(u+sb)^{\m} \d(\hp\cdot (u+sb)) f_s(\hp\cdot u) e^{\frac{p_T^2}{B}}\no
  =&(u+sb)^{\m} \d(p\cdot (u+sb)) f_s(p\cdot u) e^{\frac{p_T^2}{B}}\equiv(u+sb)^{\m}j_s\,,
\end{align}
where $\hp^\m$ is the (1+1)D momentum and $p_T^\m$ is the rest part transverse to both $u^\m$ and $b^\m$. They are defined by the (1+1)D metric $\gh^{\m\n}= u^{\m}u^{\n}-b^{\m}b^{\n}$ and transverse projector $P^{\m\n}\equiv \gh^{\m\n}-g^{\m\n}$ as $\hp^\m\equiv \gh^{\m\n}p_\n=(p\cdot u)u^\m-(p\cdot b)b^\m$ and $p_T^\m \equiv -P^{\m\n}p_\n=p^\m-\hp^\m$. These projections confine the LLL background currents to be effective (1+1)D since the $p_T$ dependent factor $e^{\frac{p_T^2}{B}}$ only gives Landau-level degeneracy $\frac{B}{2\p}$. 
The distribution function involving fermion energy is defined as
\begin{align}\label{distr}
  f_s(p\cdot u)=\frac{2}{(2\p)^3}\sum_{r=\pm}\frac{ r \th(rp\cdot u)}{e^{r \(p\cdot u-\m_s\)/T}+1}\,,
\end{align}
where $\m_s=\m+s\m_5$ is the chemical potential for chiral fermions and $T$ is temperature, which are taken as constant for simplicity. 
Also, to be concise, when solving the CKE, we focus on the right-handed fermions and suppress helicity $s$ in the notation, which will be restored in the final solution.

\section{Solution to chiral kinetic equations}\label{sec_solution}

To begin with, we examine the structure of the CKE by analyzing perturbative solution at $O(a)$ and $O(\pd a)$ separately. 
Substituting the perturbative expansion \eqref{expd} into CKE \eqref{eom1} to \eqref{eom3} and noting the background solution \eqref{cov_sol} automatically satisfies the CKE, at $O(a)$ we obtain
\begin{align}
  p_{\m} j_{a}^{\m}&=0\,,\label{eom1_0}\\
  F_{\m\n}\frac{\pd}{\pd p_\n} j_{a}^{\m}&=0\,,\label{eom2_0}\\
  p^{[\m} j_{a}^{\n]} -\frac{1}{2} \e^{\m\n\r\s} F_{\r\l}\frac{\pd}{\pd p_\l} j_{\s}^{a}&=0.\label{eom3_0}
\end{align}
Since \eqref{eom1_0} through \eqref{eom3_0} are homogeneous equations, which are generally satisfied by the solution with a structure similar to the LL states \eqref{cov_sol}, we take the ansatz $j_{a}^\m=(u+b)^\m j_a$ with $j_a\propto \d(p\cdot (u+b)) e^{\frac{p_T^2}{B}}$ as $O(a)$ redistribution of the LLL state.

At $O(\pd a)$, the CKE with constant fluid velocity can be grouped into four cases for $\cA=E_\pr,E_\pp,\cB_\pr,\cB_\pp$,
\begin{align}
  p_\m j_{\cA }^\m=&0\,,\label{eom1E} \\
  F_{\m\n}\frac{\pd}{\pd p_\n} j_{\cA }^\m=&\pd_\m^\cA j_{a}^{\m}-f_{\m\n}^{\cA}\frac{\pd}{\pd p_\n}j_{(0)}^\m\,,\label{eom2E} \\
  p^{[\m} j_{\cA }^{\n]}-\frac{1}{2}\e^{\m\n\r\s} F_{\r\l}\frac{\pd}{\pd p_\l} j_\s^{\cA }=&-\frac{1}{2}\e^{\m\n\r\s} \(\pd_\r^\cA j_{\s}^{a}-f_{\r\l}^{\cA}\frac{\pd}{\pd p_\l} j_\s^{(0)}\)\,,\label{eom3E}
\end{align}
where $\pd_\m^\cA j_a$ are defined by splitting the gradient of $j_a$ into four terms in accordance to $\cA$,
\begin{align}\label{sp_pd}
  \pd_\m j_a=&\(u_\m\bar\pd_0+w_\m^k \bar\pd_k+b_\m\bar\pd_3\)\(\ab_0\frac{\pd j_a}{\pd \ab_0}+\ab_k\frac{\pd j_a}{\pd \ab_k}+\ab_3\frac{\pd j_a}{\pd \ab_3}\)\no
  =&\(u_\m\bar\pd_0+b_\m\bar\pd_3\)\(\ab_0\frac{\pd j_a}{\pd \ab_0}+\ab_3\frac{\pd j_a}{\pd \ab_3}\)+w_\m^k \bar\pd_k \ab_i\frac{\pd j_a}{\pd \ab_i}\no
  &\quad+\(u_\m\bar\pd_0 \ab_k\frac{\pd j_a}{\pd \ab_k}+w_\m^k \bar\pd_k \ab_0\frac{\pd j_a}{\pd \ab_0}\)+\(w_\m^k \bar\pd_k \ab_3\frac{\pd j_a}{\pd \ab_3}+b_\m\bar\pd_3 \ab_k\frac{\pd j_a}{\pd \ab_k}\)\no
  \equiv&\pd_\m^{E\pr}j_a+\pd_\m^{\cB\pr}j_a+\pd_\m^{E\pp}j_a+\pd_\m^{\cB\pp}j_a\,.
\end{align}

\subsection{Perturbative solutions in all possible cases}

We start with the case of parallel electric field $E_\pr^\m$ sourced from $\pd_\m^{E\pr}j_a$. In collisonless limit, there are no dissipative transports such as electric current along $E_\m$. Therefore we exclude dissipative solution proportional to $p^\m E_\m$. Since there is no other solution at $O(\pd a)$ in the $E_\pr$ situation, we expect parallel electric field only leads to classical motion of Landau level states in the longitudinal direction. In the LLLA, this classical longitudinal motion amounts to the redistribution of the LLL state $j_{a}^\m$, which satisfies homogeneous CKE \eqref{eom1_0} to \eqref{eom3_0} automatically. In order to let the CKE \eqref{eom1E} to \eqref{eom3E} hold for a vanishing $j_{E\pr}^{\m}$, we require the inhomogeneous terms of the CKE to vanish,
\begin{align}
  \pd_\m^{E\pr} j_{a}^{\m}=&f_{\m\n}^{E\pr}\frac{\pd}{\pd p_\n}j_{(0)}^\m\,,\label{src1}\\
  \e^{\m\n\r\s} \pd_\r^{E\pr} j_{\s}^{a}=&\e^{\m\n\r\s} f_{\r\l}^{E\pr}\frac{\pd}{\pd p_\l} j_\s^{(0)}.\label{src2}
\end{align}
The above two equations give the following constraint on $j_a$,
\begin{align}\label{cj_epr}
  \ab_0\frac{\pd j_a}{\pd \ab_0}+\ab_3\frac{\pd j_a}{\pd \ab_3}=&\frac{\qb_0\ab_3-\qb_3\ab_0}{\qb_0-\qb_3} \(\frac{\pd j}{\pd\pb_0}+\frac{\pd j}{\pd\pb_3}\)\no
  =&\frac{\qb_0\ab_3-\qb_3\ab_0}{\qb_0-\qb_3} \d(\pb_0-\pb_3) f'(\pb_0) e^{\frac{p_T^2}{B}}\,.
\end{align}

We turn to the case of parallel magnetic field $\cB_\pr^\m$ induced by $\pd_\m^{\cB\pr}j_a$. One finds that $\cB_\pr^\m=\cB_\pr b^\m$ just enhances the magnitude of the longitudinal magnetic field. By the replacement $B\to B+\cB_\pr$ in \eqref{cov_sol} to $O(\pd a)$, one gets
\begin{align}\label{cb_pr_0}
  j_{(0)}^{\m}+j_{\cB\pr}^{\m}=(u+b)^{\m} \d(p\cdot (u+b)) f(p\cdot u) e^{\frac{p_T^2}{B+\cB_\pr}}\,.
\end{align}
When including $f_{\m\n}^{\cB\pr}$ into $F_{\m\n}$, \eqref{cb_pr_0} automatically satisfies the homogeneous CKE \eqref{eom1_0} to \eqref{eom3_0}, which gives 
\begin{align}\label{cb_pr}
  j_{\cB\pr}^{\m}=-(u+b)^{\m} \d(p\cdot (u+b)) f(p\cdot u) e^{\frac{p_T^2}{B}}\frac{p_T^2 \cB_\pr}{B^2}=-(u+b)^{\m}\frac{\cB_\pr}{B}\frac{p_T^2}{B} j\,.
\end{align}
This indicates \eqref{cb_pr} satisfies the CKE \eqref{eom1E} to \eqref{eom3E} by requiring vanishing gradient terms, i.e., $\pd_\m^{\cB\pr} j_a=0$. Thus we get constraints on $j_a$, 
\begin{align}
  \pd_\m^{\cB\pr} j_{a}^{\m}=0\quad\to\quad (u+b)^\m q_\m^T \ab_k \frac{\pd j_a}{\pd \ab_k}=0\,,\label{src1_cb}\\
  \e^{\m\n\r\s} \pd_\r^{\cB\pr} j_{\s}^{a}=0\quad\to\quad \e^{\m\n\r\s} (u+b)_{\s} q_\r^T \ab_k \frac{\pd j_a}{\pd \ab_k}=0.\label{src2_cb}
\end{align}
One finds \eqref{src1_cb} holds trivially while \eqref{src2_cb} gives
\begin{align}\label{cj_b_pr}
  \e^{\m\n\r\s} (u+b)_{\s} \qb_i w_\r^i \ab_k \frac{\pd j_a}{\pd \ab_k}=0 \quad\to\quad \qb_i\ab_k \frac{\pd j_a}{\pd \ab_k}=0\,.
\end{align}
This is the constraint on $j_a$ in $\cB_\pr$ case. 

The perpendicular electric field $E_\pp^\m$ is induced by $\pd_\m^{E\pp}j_a$. In collisionless limit, the chiral medium is driven into a new state, where the plasma drifts in the direction transverse to both $E_\pp^\m$ and $B^\m$. We take the following ansatz at $O(\pd a)$ for the drift state solution, 
\begin{align}\label{anz_drift}
  j_{E\pp}^{\m }=&(u+b)^{\m }\,p\cdot u_{\pp} \(\frac{\pd }{\pd(p\cdot u)}G_1+G_2\)+u_{\pp} ^{\m } G_3\,,
\end{align}
where drift velocity $u_{\pp}^{\m}=\frac{U^\m}{B}$ and $U^\m\equiv\frac{1}{2}\e^{\m \n \r \s } b_\n f_{\r\s}^{E\pp}$ with $u_\pp^\m$ orthogonal to $E_\m^{\pp},\,b_\m$ and $u_\m$.
As is shown in Appendix \eqref{app_solution}, by solving for the undetermined functions $G_n\propto \d(p\cdot(u+b))e^{ \frac{p_T^2}{B} }$, we obtain the drift solution for right-handed fermions as 
\begin{align}
  j_{E\pp}^{\m }=&(u+b)^{\m }p\cdot u_{\pp} \(\frac{\pd j}{\pd(p\cdot u)}-\frac{2p\cdot u}{B}j-\frac{\pd j_a}{\pd \ab_0}\)+u_{\pp} ^\m j.\label{drift_sol}
\end{align}
If the drift state is in equilibrium, we can neglect the term $\frac{\pd j_a}{\pd \ab_0}$ sourced by the redistribution. Then the drift solution can be combined with the zeroth order solution into a more suggestive form,
\begin{align}\label{drift_sol_compact}
  j_{(0)}^{\m }+j_{E\pp}^{\m }=&(\tu+b)^{\m }\d(p\cdot(\tu+b))f(p\cdot \tu)e^{(p^2-(p\cdot \tu)^2+(p\cdot b)^2)/B}\,.
\end{align}
This is nothing but the zeroth order solution with $u^\m\to \tu^\m\equiv(u+u_{\pp} )^\m$. This means the drift state is equivalent to the boosted equilibrium state.

In the case of the perpendicular magnetic field $\cB_\pp^\m$ caused by $\pd_\m^{\cB\pp} j_a$, the magnetic field will tilt a bit upon adding a perturbative transverse $\cB_\pp^\m$ into the background $B^\m$. The analysis is similar to the $E_\pp$ case. Taking the ansatz similar to \eqref{anz_drift},
\begin{align}\label{anz_tilt}
  j_{\cB\pp}^{\m }=&(u+b)^{\m }\,\frac{p\cdot \cB_\pp}{B}\(\frac{\pd }{\pd(p\cdot u)}H_1+H_2\)+\frac{\cB_\pp^{\m }}{B} H_3\,,
\end{align}
one gets the tilted solution for right-handed fermions as
\begin{align}
  j_{\cB\pp}^{\m }=&(u+b)^{\m }\frac{p\cdot \cB_\pp}{B}\(\frac{\pd j}{\pd(p\cdot b)}-\frac{2p\cdot u}{B}j+\frac{\pd j_a}{\pd \ab_3}\)+\frac{\cB_\pp^\m}{B} j.\label{tilt_sol}
\end{align}
Again, by dropping the redistribution mode $\frac{j_a}{\pd \ab_3}$, the tilted solution can be connected with the zeroth order solution into
\begin{align}\label{tilt_sol_compact}
  j_{(0)}^{\m }+j_{\cB\pp}^{\m }=&(u+\tb)^{\m }\d(p\cdot(u+\tb))f(p\cdot u)e^{(p^2-(p\cdot u)^2+(p\cdot \tb)^2)/B}\,.
\end{align}
This is just the zeroth order solution with $b^\m\to \tb^\m\equiv b^\m+\cB_\pp^\m/B$. Thus we can understand the tilted state as an equivalent to rotated equilibrium state.

At last, in arriving at the $O(\pd a)$ drift solution \eqref{drift_sol} and tilted solution \eqref{tilt_sol}, there are constraints on the $O(a)$ source $j_a$,
\begin{align}\label{cj_pp}
  \frac{\pd j_a}{\pd \ab_k}\ab_k\(\qb_0-\qb_3\)=0\,,\qquad \(\frac{\pd j_a}{\pd \ab_0}\qb_0+\frac{\pd j_a}{\pd \ab_3}\qb_3\)\ab_k=0\,.
\end{align}
Combining the constraints on $j_a$ in \eqref{cj_epr}, \eqref{cj_b_pr} and \eqref{cj_pp}, we obtain the $O(a)$ redistribution of the LLL state $j_{a}^\m$ as,
\begin{align}\label{ja_sol}
  j_{a}^\m=(u+b)^\m\frac{\qb_0\ab_3-\qb_3\ab_0}{\qb_0-\qb_3} \d(\pb_0-\pb_3) f'(\pb_0) e^{\frac{p_T^2}{B}}\,.
\end{align}
Here we correct a minor mistake in the drift and tilted solution of \cite{Gao:2020ksg} where the medium dependent terms $\frac{2p\cdot u}{B}j$ and $\frac{2p\cdot b}{B}j$ should be attributed to the $O(\pd a)$ solutions \eqref{drift_sol} and \eqref{tilt_sol} instead of the $O(a)$ solution \eqref{ja_sol}. We will see in the following subsections that only in this way will the currents satisfy the covariant anomalies \eqref{ward_kd_2} and the stress tensor match with MHD in the limit $|\qb_3|\ll \qb_0\to0$. Also, the resultant correlators will meet derivative symmetry relations and agree with the result in field theory approach \cite{PhysRevD.83.111501}.

\subsection{Final solution}

Restoring helicity $s=\pm$ for right/left-handed fermions, we get
\begin{align}
  j_{(0) s}^\m=&\, (u+sb)^\m\d(\pb_0-s\pb_3) f_s(\pb_0) e^{\frac{p_T^2}{B}}\equiv(u+sb)^\m j_s\,,\no
  j_{a s}^\m=&\, (u+sb)^\m\frac{\qb_0\ab_3^s-\qb_3\ab_0^s}{s\qb_0-\qb_3} \d(\pb_0-s\pb_3) f_s'(\pb_0) e^{\frac{p_T^2}{B}}\,,\no
  j_{E\pr s}^{\m}=&\, 0\,,\no
  j_{\cB\pr s}^{\m}=&\, -(u+sb)^{\m}\frac{\cB_{\pr s}}{B}\frac{p_T^2}{B} j_s\,,\no
  j_{E\pp s}^{\m }=&\, (u+sb)^{\m }\frac{p\cdot U_s}{B} \(\frac{\pd j_s}{\pd(p\cdot u)}-\frac{2p\cdot u}{B}j_s-\frac{\pd j_{a s}}{\pd \ab_0^s}\)+\frac{U_s^\m}{B} j_s\,,\no
  j_{\cB\pp s}^{\m }=&\, (u+sb)^{\m }\frac{p\cdot \cB_{\pp s}}{B}\(\frac{\pd j_s}{\pd(p\cdot b)}+\frac{2p\cdot b}{B}j_s+\frac{\pd j_{a s}}{\pd \ab_3^s}\)+s\frac{\cB_{\pp s}^\m}{B} j_s\,,
\end{align}
where $\cB_s^\m=\cB^\m+s\cB_5^\m$ and $U_s^{\m}=U^\m+sU_5^\m$ are defined by $f_{\m\n}^s=f_{\m\n}+sf_{\m\n}^5$. 
After integration over momenta and summation/subtraction over right/left-handed contributions, we get the constitutive relation of currents,
\begin{align}
  J_{(0)}^{\m}=&\frac{\m B}{2\p^2}u^\m+\frac{\m_5 B}{2\p^2}b^\m\,,\label{j0}\\
  J_{(0)5}^{\m}=&\frac{\m_5 B}{2\p^2} u^\m+\frac{\m B}{2\p^2}b^\m\,,\label{j05}\\
  J_{a}^{\m}=&\frac{\D\m B}{2\p^2}u^\m+\frac{\D\m_5 B}{2\p^2}b^\m\,,\label{ja}\\
  J_{a5}^{\m}=&\frac{\D\m_5 B}{2\p^2} u^\m+\frac{\D\m B}{2\p^2}b^\m\,,\label{ja5}\\
  J_{\cA }^{\m}=&\frac{\m \(U^\m+\cB_\pr u^\m\) + \m_5 \(U_5^\m+\cB_{\pr 5}u^\m\)}{2\p^2}+\frac{\m_5\(\cB_\pp^\m+\cB_\pr^\m\)+\m\(\cB_{\pp 5}^\m+\cB_{\pr 5}^\m\)}{2\p^2}\,,\label{j1}\\
  J_{\cA 5}^{\m}=&\frac{\m_5 \(U^\m+\cB_\pr u^\m\) + \m \(U_5^\m+\cB_{\pr 5}u^\m\)}{2\p^2}+\frac{\m\(\cB_\pp^\m+\cB_\pr^\m\)+\m_5\(\cB_{\pp 5}^\m+\cB_{\pr 5}^\m\)}{2\p^2}\,.\label{j15}
\end{align}
The $O(a)$ currents \eqref{ja} and \eqref{ja5} as the redistribution of the LLL state, indeed reproduce the density waves \eqref{cmw} which are dictated by covariant anomalies \eqref{ward_kd_2}. This confirms that the gauge invariant currents solved from CKT are the covariant ones. One readily gets the consistent currents by adding the Chern-Simons current according to \eqref{cs_cv_2}, where it turns out that only $J_a^\m$ is changed to $\cJ_a^\m$,
\begin{align}\label{cs_cv_0a}
  \cJ_{a}^\m=&\frac{ B}{2\p^2}\qb_3\frac{\qb_3\ab_0-\qb_0\ab_3+\qb_0\ab_0^5-\qb_3\ab_3^5}{\qb_0^2-\qb_3^2}u^\m\no
  &+\frac{ B}{2\p^2}\qb_0\frac{\qb_3\ab_0-\qb_0\ab_3+\qb_0\ab_0^5-\qb_3\ab_3^5}{\qb_0^2-\qb_3^2}b^\m\,.
\end{align}
One also finds the currents \eqref{ja} and \eqref{ja5} satisfy the (1+1)D constitutive relation \eqref{j5_j}. 
The $O(\pd a)$ currents \eqref{j1} and \eqref{j15} go beyond (1+1)D in the transverse direction due to the perpendicular gradients of the gauge fields. 
The terms in the direction of $U^\m$ in \eqref{j1} and \eqref{j15} are the vector and axial Hall currents which are transverse to $E_{\pp}^\m$, $b^\m$ and $u^\m$.
The parts along the magnetic field in \eqref{j0}, \eqref{ja} and \eqref{j1} give the well-known CME at $O(1)$, $O(a)$ and $O(\pd a)$ separately, while those in \eqref{j05}, \eqref{ja5} and \eqref{j15} give CSE. 

The constitutive relation of stress tensor writes
\begin{align}
  T_{(0)}^{\m\n}=&\frac{\c B}{2\p^2}(u^\m u^\n+b^\m b^\n)+\frac{\c_5 B}{2\p^2}u^{\{\m }b^{\n\}}\,,\label{T0}\\
  T_{a}^{\m\n}=&\frac{\(\m\D\m+\m_5\D\m_5\)B}{2\p^2}(u^\m u^\n+b^\m b^\n)+\frac{\(\m\D\m_5+\m_5\D\m\)B}{2\p^2}u^{\{\m }b^{\n\}}\,,\label{Ta0}\\
  T_{\cA }^{\m\n}=&\frac{\c \cB_\pr + \c_5 \cB_{\pr 5}}{2\p^2}(u^\m u^\n+b^\m b^\n)+\frac{\c_5 \cB_\pr + \c \cB_{\pr 5}}{2\p^2}u^{\{\m }b^{\n\}}\no
  &+\frac{1}{2\p^2}\(\frac{\qb_0^2}{\qb_0^2-\qb_3^2}\frac{B}{4}+\c\)\(u^{\{\m}U^{\n\}}+u^{\{\m}\cB_{\pp 5}^{\n\}}+U_5^{\{\m }b^{\n\}}+\cB_{\pp}^{\{\m }b^{\n\}}\)\no
  &+\frac{1}{2\p^2}\(\frac{\qb_0\qb_3}{\qb_0^2-\qb_3^2}\frac{B}{4}+\c_5\)\(u^{\{\m}U_5^{\n\}}+u^{\{\m}\cB_\pp^{\n\}}+U^{\{\m }b^{\n\}}+\cB_{\pp\ 5}^{\{\m }b^{\n\}}\)\,,\label{T1}
\end{align}
where $\c=\frac{\m^2+\m_5^2}{2}+\frac{\p^2T^2}{6}$ and $\c_5=\m\m_5$. There are waves in stress tensor at $O(a)$ and $O(\pd a)$ accompanying the density waves of vector and axial charges. In \eqref{Ta0}, the $O(a)$ terms $u^\m u^\n$, $u^{\{\m} b^{\n\}}$ and $b^\m b^\n$ can be identified as the waves of energy density, longitudinal heat flow and longitudinal momentem flow. Beyond (1+1)D, there are $O(\pd a)$ waves $\frac{\qb_0^2}{\qb_0^2-\qb_3^2}\frac{B}{4}$ and $\frac{\qb_0\qb_3}{\qb_0^2-\qb_3^2}\frac{B}{4}$ in the transverse heat flow and momentem flow of \eqref{T1}, which are caused by the perpendicular gradients of the redistribution $J_a^\m$. Interestingly, the redistribution does not induce waves in the $O(\pd a)$ currents \eqref{j1} and \eqref{j15} since the wave dependent terms in the solution are odd functions of transverse momentem $p_T$ and therefore vanishing upon momenta integration.

\subsection{Correlators and effective action}

After the functional differentiation performed in Appendix \eqref{app_functional}, we obtain the current-current correlators as follows. 
For the $O(a)$ currents, we have
\begin{align}\label{grn_0}
  G_{VV,a}^{\m\l}(q)=G_{AA,a}^{\m\l}(q)=&\frac{B}{2\p^2}\frac{\qb_3^2 u^\m u^\l+\qb_0^2 b^\m b^\l+\qb_0\qb_3 u^{\{\m} b^{\l\}}}{\qb_0^2-\qb_3^2}\,,\no
  G_{VA,a}^{\m\l}(q)=G_{AV,a}^{\m\l}(q)=&\frac{B}{2\p^2}\frac{\qb_0^2 u^\m b^\l+\qb_3^2 b^\m u^\l+\qb_0\qb_3\(u^\m u^\l+b^\m b^\l\)}{\qb_0^2-\qb_3^2}\,.
\end{align}
Noting $\gh^{\m\n}=u^\m u^\n-b^\m b^\n$ and $\qh^\m=\qb_0 u^\m+\qb_3 b^\m$ in (1+1)D, one finds that $G_{VV,a}$ agrees with \cite{PhysRevD.83.111501}. 
Beyond (1+1)D, the correlators for $O(\pd a)$ currents are
\begin{align}\label{grn_1}
  G_{VV,\cA}^{\m\l}(q)=&\frac{i\,\m_5 }{2\p^2}\(\e^{\m \l \r \s } \qb_3 - b^{[\m}\e^{\l] \n \r \s }q_\n^T\)u_\r b_\s-\frac{i\,\m }{2\p^2}\(\e^{\m \l \r \s }\qb_0 + u^{[\m}\e^{\l] \n \r \s }q_\n^T\)u_\r b_\s\,,\no
  G_{VA,\cA}^{\m\l}(q)=&\frac{i\,\m }{2\p^2}\(\e^{\m \l \r \s } \qb_3 - b^{[\m}\e^{\l] \n \r \s }q_\n^T\)u_\r b_\s-\frac{i\,\m_5 }{2\p^2}\(\e^{\m \l \r \s }\qb_0 + u^{[\m}\e^{\l] \n \r \s }q_\n^T\)u_\r b_\s\,,
\end{align}
where one also has $G_{VV,\cA}^{\m\l}(q)=G_{AA,\cA}^{\m\l}(q)$ and $G_{VA,\cA}^{\m\l}(q)=G_{AV,\cA}^{\m\l}(q)$. 
Using the relations \eqref{cor_cons_cov}, one can easily get the consistent correlators $\cG$ from the covariant ones $G$. The only difference is in $\cG_{VA,a}$ where the addition of the Chern-Simons current gives,
\begin{align}
  \cG_{VA,a}^{\m\l}(q)=\frac{B}{2\p^2}\frac{\qb_3^2 u^\m b^\l+\qb_0^2 b^\m u^\l+\qb_0\qb_3\(u^\m u^\l+b^\m b^\l\)}{\qb_0^2-\qb_3^2}\,.
\end{align}
The resulting consistent correlators $\cG=\cG_a+\cG_{\cA }$ perfectly fit the derivative symmetry relations \eqref{bose_sym}. 
This allows us to construct the effective action for the perturbtive gauge fields using \eqref{cor_cons}. For the $O(a)$ perturbations, one gets
\begin{align}\label{lag_0}
  \cL_{a}
  =&\frac{B}{8\p^2}\frac{1}{\hat k^2}\[\(f_{E\pr}^{\m\n}(k) + 2i\hat k^\r a_\r^5(k)\e^{\m\n}\)f_{\m\n}^{E\pr}(-k) + f_{E\pr5}^{\m\n}(k)f_{\m\n}^{E\pr5}(-k)\]\,,
\end{align}
where the Fourier transformed $f_{\r\s}^{E\pr}(k)$ and $f_{\r\s}^{E\pr5}(k)$ are defined in Appendix \eqref{app_EMfield}. It can be understood as the effective Lagrangian density of perturbative gauge fields in (1+1)D theory, where apart from the free terms $\(f_{E\pr}\)^2$ and $\(f_{E\pr5}\)^2$ for gauge fields, there is an anomalous term $\hat \pd\cdot a_5\,|f_{E\pr}|$ for the consistent anomaly  \eqref{ward_kd_2}. The density waves of charges are characterized by the non-locality $1/\hat k^2$, which together with the Landau-level degeneracy $\frac{B}{2\p}$ gives a dimensionless factor in \eqref{lag_0}. 
For $O(\pd a)$ perturbations beyond (1+1)D, we get
\begin{align}\label{lag_1}
  \cL_{\cA }
  =\frac{1}{4\p^2}\e^{\l\n\r\s} \Big[\(f_{\r\s}(k) a_\l(-k) + f_{\r\s}^5(k) a_\l^5(-k)\)\(\m b_\n + \m_5 u_\n\)\no
   + \(f_{\r\s}(k) a_\l^5(-k) + f_{\r\s}^5(k) a_\l(-k)\) \(\m_5\,b_\n + \m\,u_\n\)\Big]\,.
\end{align}
The above terms do not contribute to chiral anomaly. If we treat $\hat A_\n\equiv \m b_\n + \m_5 u_\n$ and $\hat A_\n^5\equiv \m_5\,b_\n + \m\,u_\n= -\e_{\n\m}\hat A^\m$ as effective vector and axial gauge potentials of the chiral medium, the terms in \eqref{lag_1} can be reckoned as Chern-Simons structures, like $\e^{\l\n\r\s}a_\l \hat A_\n \pd_\r a_\s$, etc. The effective Lagrangian density $\cL(a,a_5,k)=\cL_{a}+\cL_{\cA }$, as a generating functional, captures all the necessary information to regains the consistent currents $\cJ$ and correlators $\cG$ in momentum space by noting \eqref{cor_cons} and
\begin{align}
  \int& d^4x\, a^\m(x) \cJ_\m(x)=\int \frac{d^4k}{(2\p)^4} a^\m(k) \cJ_\m(-k)\no
  \to&\qquad\frac{\d \cL}{\d a_\n(q)}=\cJ^\n(-q)\,,\qquad \frac{\d \cL}{\d a_\n(-q)}=\cJ^\n(q)\,.
\end{align}
The effective action as volume integral of $\cL$ writes
\begin{align}
  \G(a,a_5)=\int \frac{d^4k}{(2\p)^4}\cL(a,a_5,k)=\int d^4x\, \cL(a,a_5,x)\,,
\end{align}
where $\cL(a,a_5,x)$ is the effective Lagrangian density in coordinate space. Noting the relations
\begin{align}
  \int d^4x\,  a_\l(x)f_{\r\s}(x)=&\int \frac{d^4k}{(2\p)^4} a_\l(-k)f_{\r\s}(k)\,,\no
  \int d^4x\,d^4y\, \cD(x-y)f^{\m\n}(x)f_{\m\n}(y)=&\int \frac{d^4k}{(2\p)^4}\frac{1}{\hat k^2}f^{\m\n}(-k)f_{\m\n}(k)\,,\no
  \int d^4x\,d^4y\, \cD(x-y)\hat \pd^\r a_\r^5(x)f_{\m\n}(y)=&\int \frac{d^4k}{(2\p)^4}\frac{1}{\hat k^2}i\hat k^\r a_\r^5(-k)f_{\m\n}(k)\,,
\end{align}
we can write the effective action as
\begin{align}
  \G=\frac{B}{8\p^2}\int d^4x d^4y\cD(x-y)\Big[\(f_{E\pr}^{\m\n}(x)+2\hat \pd^\r a_\r^5(x)\e^{\m\n}\)f_{\m\n}^{E\pr}(y)+f_{E\pr5}^{\m\n}(x)f_{\m\n}^{E\pr5}(y)\Big]\no
  +\frac{1}{4\p^2}\int d^4x\e^{\l\n\r\s} \Big[\(f_{\r\s}(x) a_\l(x) + f_{\r\s}^5(x) a_\l^5(x)\)\(\m b_\n + \m_5 u_\n\)\no
   + \(f_{\r\s}(x) a_\l^5(x) + f_{\r\s}^5(x) a_\l(x)\) \(\m_5\,b_\n + \m\,u_\n\)\Big]\,.
\end{align}

Additionally, the correlators between stress tensor and currents also satisfy the derivative symmetry relations, 
\begin{align}\label{TJ_sym}
  \cG_{TV}^{\l\n,\m}(q)=\cG_{VT}^{\m,\l\n}(-q)\,,\qquad
  \cG_{TA}^{\l\n,\m}(q)=\cG_{AT}^{\m,\l\n}(-q)\,,
\end{align}
where the correlators are defined as
\begin{align}\label{TJ_cor}
  \cG_{TV}^{\l\n,\m}\equiv& \frac{\d T_{\cA }^{\l\n}}{\d a_\m}\,,\qquad
  \cG_{VT}^{\m,\l\n}\equiv 2\frac{\d J_{\cV}^\m}{\d g_{\l\n}}\,,\no
  \cG_{TA}^{\l\n,\m}\equiv& \frac{\d T_{\cA }^{\l\n}}{\d a_\m^5}\,,\qquad
  \cG_{AT}^{\m,\l\n}\equiv 2\frac{\d J_{\cV5}^\m}{\d g_{\l\n}}\,,
\end{align}
with the vortical solution derived in \cite{Lin:2021sjw},
\begin{align}\label{j_vor}
  J_{\cV }^\m=&\frac{\o}{2\p^2}\(2\c+\frac{B}{2}\)u^\m+\frac{\o}{2\p^2}2\c_5 b^\m\,,\no
  J_{\cV 5}^\m=&\frac{\o}{2\p^2}2\c_5 u^\m+\frac{\o}{2\p^2}\(2\c+\frac{B}{2}\) b^\m\,.
\end{align}
We have assumed the vortical solution is immune to the distinction of consistent and covariant anomalies since the influence has been covered in the gauge field solution. The vector and axial currents in \eqref{j_vor} are also related by the leading order constitutive relation \eqref{j5_j}. The former implies the generation of charge density and longitudinal CVE current driven by a parallel vorticity, which is respectively reciprocal to that of the transverse heat flows along the Hall and CME currents induced by the perpendicular electromagnetic field. The latter as axial counterpart is similar. We then extract the relevant reciprocals from the stress tensor \eqref{T1},
\begin{align}\label{T1_nw}
  T_{\cA }^{\m\n}=&\frac{1}{2\p^2}\(\frac{B}{4}+\c \)u^{\{\m}U^{\n\}}+\frac{\c_5}{2\p^2} u^{\{\m}\cB_\pp^{\n\}}\no
  &+\frac{\c_5}{2\p^2} u^{\{\m}U_5^{\n\}}+\frac{1}{2\p^2}\(\frac{B}{4}+\c\)u^{\{\m}\cB_{\pp 5}^{\n\}}\,,
\end{align} 
where we have taken the long wavelength limit $\qb_3\to0$ before the static limit $\qb_0\to0$. The variation of $T_{\cA }$ \eqref{T1_nw} is similar to the case of $J_{\cA }$ \eqref{j1}. Using \eqref{TJ_cor}, one gets
\begin{align}
  \cG_{TV}^{\l\n,\m}(q)=\frac{1}{2\p^2}\(\frac{B}{4}+\c \)u^{\{\l}\e^{\n\} \a \r \s } u^\m iq_\a^T u_\r b_\s+\frac{\c_5}{2\p^2}u^{\{\l}\e^{\n\} \a \r \s } b^\m iq_\a^T u_\r b_\s\,,\no
  \cG_{TA}^{\l\n,\m}(q)=\frac{\c_5}{2\p^2}u^{\{\l}\e^{\n\} \a \r \s } u^\m iq_\a^T u_\r b_\s+\frac{1}{2\p^2}\(\frac{B}{4}+\c\)u^{\{\l}\e^{\n\} \a \r \s } b^\m iq_\a^T u_\r b_\s\,.
\end{align}
In order to calculate $\cG_{VT}^{\m,\l\n}(-q)$ and $\cG_{AT}^{\m,\l\n}(-q)$ by the variation of \eqref{j_vor} with respect to $\d g_{\l\n}$, we note that the longitudinal vorticity $\o$ comes from the tranverse gradients of the metric, i.e., $-\o=b_\s \o^\s=\frac{1}{2}\e^{\s\r\a\b}b_\s u_\r \pd_\a^T \d u_\b\to\frac{1}{2}\e^{\s\r\a\b}b_\s u_\r ik_\a^T \d u_\b(k)$ with $\d u_\b(k)=u^\g \d g_{\g\b}(k)$, and the $\l\n$ symmetrized variation writes,
\begin{align}
  \frac{\d g_{\g\b}(k)}{\d g_{\l\n}(q)}=\int d^4k \frac{1}{2}\(\d_\g^\l\d_\b^\n+\d_\g^\n\d_\b^\l\)\d^{(4)}(k-q)\,.
\end{align} 
Then one easily verifies the derivative symmetry relations \eqref{TJ_sym}. In principle, one can also construct the effective action with the coupling of the perturbative metric and gauge fields by repeating the generating functional procedure. However, we note the derivative symmetry is only fulfilled in the limit $|\qb_3|\ll \qb_0\to0$, and the metric is designed for the longitudinal vorticity, which is a rather restricted situation. The effective action in such case does not have a general meaning. 

Finally, we verify the two-point functions meet the Onsager relations,
\begin{align}
  \cG_{ab}(q_0,{\bf q},u_0,{\bf u},\tilde B_0,\tilde {\bf B})=\g_a\g_b\cG_{ba}(q_0,-{\bf q},u_0,-{\bf u},\tilde B_0,-\tilde{\bf B})\,,
\end{align} 
where $\g=\pm1$ is the time-reversal eigenvalue of the operator and $a\,,b$ are the two operators in the correlator. The signs flip under time-reversal in the spatial components of momentum ${\bf q}$, fluid velocity ${\bf u}$ and magnetic field $\tilde {\bf B}={\bf B}+\cB$ (or its longitudinal ${\bf b}$ and transverse ${\bf w}_{n=1,2}$ directions). In the local rest frame with $u^\m$, $w_n^\m$ and $b^\m$ given in Appendix \eqref{app_momenta}, one readily reduces the derivative symmetry and Onsager relations to
\begin{align}
    \cG_{ab}(q_0,{\bf q},\tilde {\bf B})=\cG_{ba}(-q_0,-{\bf q},\tilde{\bf B})=\g_a\g_b\cG_{ba}(q_0,-{\bf q},-\tilde{\bf B})\,,
\end{align} 
where, by taking the default parameters $(q_0,{\bf q},\tilde {\bf B})$, and noting the operators $J_n$, $J_3$, $J_n^5$, $J_3^5$, $T_{0n}$, $T_{03}$ have $\g=-1$ while $J_0$, $J_0^5$, $T_{00}$, $T_{nn}$, $T_{33}$, $T_{nm}$, $T_{n3}$ have $\g=1$, the non-trivial correlators write
\begin{align}
  &\cG_{[J_0 J_0]}=\cG_{[J_0^5 J_0^5]}=\cG_{[J_0 J_3^5]}=\frac{B}{2\p^2}\frac{q_3^2}{q_0^2-q_3^2}\,,\quad
  \cG_{[J_3 J_3]}=\cG_{[J_3^5 J_3^5]}=\cG_{[J_3 J_0^5]}=\frac{B}{2\p^2}\frac{q_0^2}{q_0^2-q_3^2}\,,\no
  &\cG_{[J_0 J_3]}=\cG_{[J_0^5 J_3^5]}=\cG_{[J_0 J_0^5]}=\cG_{[J_3 J_3^5]}=\frac{B}{2\p^2}\frac{q_0 q_3}{q_0^2-q_3^2}\,,\no
  &\cG_{\{J_0 J_n\}}=\cG_{\{J_0^5 J_n^5\}}=\cG_{\{J_3 J_n^5\}}=-\cG_{\{J_n J_3^5\}}=-\frac{i\m}{2\p^2}\e^{nm}q_m\,,\no
  &\cG_{\{J_3 J_n\}}=\cG_{\{J_3^5 J_n^5\}}=\cG_{\{J_0 J_n^5\}}=-\cG_{\{J_n J_0^5\}}=-\frac{i\m_5}{2\p^2}\e^{nm}q_m\,,\no
  &\cG_{\{J_n J_m\}}=\cG_{\{J_n^5 J_m^5\}}=\frac{i}{2\p^2}\e^{nm}\(\m_5 q_3-\m q_0\)\,,\no
  &\cG_{\{J_n J_m^5\}}=-\cG_{\{J_m J_n^5\}}=\frac{i}{2\p^2}\e^{nm}\(\m q_3-\m_5 q_0\)\,,\no
  &\cG_{\{T_{0n} J_0\}}=\cG_{\{T_{0n} J_3^5\}}=\frac{i}{2\p^2}\e^{nm}q_m\(\frac{B}{4}+\c\)\,,\quad
  \cG_{\{T_{0n} J_3\}}=\cG_{\{T_{0n} J_0^5\}}=\frac{i}{2\p^2}\e^{nm}q_m\c_5\,,
\end{align} 
where we have denoted $\cG_{[ab]}$ as $\cG_{ab}=\cG_{ba}$ and $\cG_{\{ab\}}$ as $\cG_{ab}=-\cG_{ba}$.

\subsection{Discussion}

We have seen the CKT result satisfies the derivative symmetry and Onsager relations by taking the limit $\qb_3\to0$ before $\qb_0\to0$. However, another limit by taking $\qb_0\to0$ before $\qb_3\to0$ gives $\D\m=-\ab_0$ and $\D\m_5=-\ab_0^5$ in \eqref{che_cmw}, which yields the static-equilbrium condition $\pd_\l\D\m=E_\l$. This means that a plasma subject to a static external electric field will develop a gradient in the chemical potential which compensates the applied field to maintain the static state. It seems that both of the two limits have physical validity in their own right, though they are non-commutable. For the former limit, one has to take an ad hoc constraint $a_5^\m=-\e^{\m\n}a_\n$ on the external gauge fields, which gives $\D\m=-\ab_3^5=-\ab_0$ and $\D\m_5=-\ab_0^5=-\ab_3$ to arrive at static-equilbrium. A similar non-commutativity was also found in \cite{Son:2012zy}, where the CKT result gives the correct conductivity of CME in the former limit. In fact, the non-commutativity is an artifact of the free fermion theory, as is pointed out in \cite{Satow:2014lva}, which should vanish with finite interactions. While the CKT with collisional term is beyond the scope of this work, we expect the Onsager relations and static-equilbrium will consist with each other in the commutable limits of $\qb_0\to0$ and $\qb_3\to0$ for an interactive theory. 

Another fact we would like to point out is that there is no general matching between the CKT result and magnetohydrodynamics (MHD) \cite{Hernandez:2017mch}. It may be understood that in MHD the conservation of axial charge suffers from QCD anomaly, in which case the CKT solution will be modified nontrivially. However, since the conservation of electric charge survives from the anomaly, the matching can be done by considering electric charge only and omitting the chiral imbalance in CKT. One then has $\m_5=0$, $a_5=0$ and $\D \m_5=0$. In the limit $|\qb_3|\ll\qb_0\to0$, $\D\m$ is also vanishing. The currents and stress tensor in CKT reduce to the simple case in \cite{Lin:2021sjw} where a consistent matching with MHD has been done. 

\section{Summary and Outlook}\label{sec_summary}

In summary, we have studied the two-point functions in the presence of perturbative vector and axial gauge fields with a strong background magnetic field. From the difference between consistent and covariant anomalies, we have obtained the relations of the correlators for consistent currents and covariant currents, which are non-trivial due to the addition of the Chern-Simons currents. The consistent correlators are constrained by the derivative symmetry while the covariant ones are gauge invariant. In the strong mangetic field limit, we have derived the density waves of electric and chiral charges which contain CMW from the effective (1+1)D anomaly equations.

We have also obtained the correlators from a Lorentz covariant CKT in the same external field configuration. The currents and stress tensor are derived by exhausting all the possible solutions to the CKE with general perturbative vector and axial gauge fields. We have written the constitutive relations of the currents and stress tensor which involve the same CMW as the above result from anomaly equations. We have confirmed that the CKT yields covariant currents, while the consistent currents have been calculated from the covariant ones by adding the Chern-Simons currents. The resulting consistent correlators satisfy the derivative symmetry and Onsager relations. The generating functional of the correlators have been constructed as the effective action for the perturbative gauge fields in the chiral plasma. 

Note that we have solved the CKT in the LLLA and collisonless limit. When the background magnectic field is not strong enough, higher Landau levels should be considered. It would be interesting to see how they will contribute to the solutions and correlators. Furthermore, for a more accurate picture of the real world with dissipation, a collisonal term should be considered, where the dissipative effects such as Ohm current along electric field will emerge and the dynamical nature of the electromagnetic field will give a more physical description of the CMW. Also, the introducing of collision and interaction may resolve the non-commutativity of static and long wavelength limits and yield a more meaningful framework of the kinetic theory. We leave it for future work.

\begin{acknowledgments}
I am grateful to the 14th workshop on QCD phase transition and relativistic heavy-ion physics for providing an stimulating environment in the final stage of this work. I also sincerely thank Prof. Shu Lin for useful discussions and collaborations on related works.
\end{acknowledgments}

\appendix

\section{Momenta Calculus}\label{app_momenta}

Upon solving the CKE, we do the momenta differentiation and integration with $\pb^\m=\L_\n^\m p^\n$ as independent variables with
\begin{align}
  \L_\n^\m=\(u^\n,\,-w_n^\n,\,-b^\n\)^T\,,\qquad
  {\L^{{}^{-1}}}_\n^{\m}\equiv\Lb_\n^\m=\(u^\m,\,w_n^\m,\,b^\m\)\,,
\end{align}
being Lorentz transformations between $p^\m$ and $\pb^\m$. Here $u^\m$ is the fluid velocity, while $w_n^\m$ (with transverse index $n=1,2$ ) and $b^\m$ are the direction vectors of transverse and longitudinal magnetic fields. One finds $\L_\m^0=\Lb_0^{\m}=u^\m,\,\L_\m^n=\Lb_n^{\m}=w_n^\m$ and $\L_\m^3=\Lb_3^{\m}=b^\m$ are basis column vectors which are orthogonal to one another and normalized as $u^2=1,\,w_n^2=b^2=-1$. The metric in the most minus signature can be written as $g^{\m\n}=u^\m u^\n-w_n^\m w_n^\n-b^\m b^\n$ with summation over transverse index $n$. Then $p^2=(p\cdot u)^2-(p\cdot w_n)(p\cdot w_n)-(p\cdot b)^2=\pb^2\equiv \pb_0^2-\pb_\pp^2-\pb_3^2$ with $\pb_0\equiv p\cdot u,\,\pb_n\equiv-p\cdot w_n,\,\pb_3\equiv-p\cdot b$, which gives $p_T^\m=\pb_n w_n^\m$ and $p_T^2=-\pb_\pp^2$. One also has $\d_\n^\m=u^\m u_\n-w_n^\m w_\n^n-b^\m b_\n$. In the local rest frame, we have $u^\m=(1,0,0,0)$, $w_1^\m=(0,1,0,0)$, $w_2^\m=(0,0,1,0)$ and $b^\m=(0,0,0,1)$. 

For more details about momenta differentiation and integration, one can look up in the appendix of \cite{Lin:2021sjw} where the component $\cT_{\cA }^{\m\n}$ of the stress tensor in the drift case is missed. We demonstrate the calculation of $\cT_{\cA }^{\m\n}$ as follows. Using $\D^{\a \b}T_{\a\b}^{E\pp}=0$, one gets
\begin{align}\label{cte}
  \cT_{E\pp}^{\m\n}&=\frac{1}{2}\(\D^{\m \a}\D^{\n \b}+\D^{\n \a}\D^{\m \b}-\frac{2}{3}\D^{\m \n}\D^{\a \b}\)T_{\a\b}^{E\pp}=\D^{\m \a}\D^{\n \b}T_{\a\b}^{E\pp}\no
  &=\frac{1}{2}\int d^4p\,\sum_{s=\pm}\[s\frac{p\cdot U_s}{B}p_T^{\{\m }b^{\n\}}\(\frac{\pd }{\pd(p\cdot u)}j_s-\frac{2p\cdot u}{B}j_s-\frac{\pd j_{a s}}{\pd \ab_0^s}\) - p\cdot b\,\frac{U_s^{\{\m }b^{\n\}}}{B}j_s\]\no
  &=\frac{1}{2}\int d^4\pb\sum_{s=\pm}\frac{U_s^{\{\m }b^{\n\}}}{B}\[-s\frac{\pb_\pp^2}{2}\(\frac{\pd }{\pd \pb_0}j_s-\frac{2\pb_0}{B}j_s-\frac{\pd j_{a s}}{\pd \ab_0^s}\) + \pb_3 j_s\]\no
  &=\frac{1}{2\p^2}\(\frac{\qb_0\qb_3}{\qb_0^2-\qb_3^2}\frac{B}{4}+\c_5\)U^{\{\m }b^{\n\}}+\frac{1}{2\p^2}\[\(1+\frac{\qb_3^2}{\qb_0^2-\qb_3^2}\)\frac{B}{4}+\c\]U_5^{\{\m }b^{\n\}}\,.
\end{align}
In the absence of charge redistribution caused by longitudinal electric field, this component is vanishing with $\m_5=0$ and $a_5=0$. Therefore the matching with MHD in \cite{Lin:2021sjw} is not affected by this missing part. The computation of another nontrivial part $\cT_{\cB\pp}^{\m\n}$ in the tilted case is similar,
\begin{align}\label{ctb}
  \cT_{\cB\pp}^{\m\n}&=\frac{1}{2}\int d^4\pb\sum_{s=\pm}\frac{\cB_{\pp\,s}^{\{\m }b^{\n\}}}{B}\[-s\frac{\pb_\pp^2}{2}\(-\frac{\pd }{\pd \pb_3}j_s-\frac{2\pb_3}{B}j_s+\frac{\pd j_{a s}}{\pd \ab_3^s}\) + s\pb_3 j_s\]\no
  &=\frac{1}{2\p^2}\(\frac{\qb_0^2}{\qb_0^2-\qb_3^2}\frac{B}{4}+\c\)\cB_{\pp}^{\{\m }b^{\n\}}+\frac{1}{2\p^2}\(\frac{\qb_0\qb_3}{\qb_0^2-\qb_3^2}\frac{B}{4}+\c_5\)\cB_{\pp\ 5}^{\{\m }b^{\n\}}\,.
\end{align}

\section{Electromagnetic Field}\label{app_EMfield}

We turn on an $O(a)$ perturbation $a^\m$ in gauge field. For constant $u$ and $b$, the perturbation $f^{\m\n}$ can also be Lorentz transformed into the local rest frame (LRF) as $\fb^{\m\n}=\pdb^{[\m} \ab^{\n]}=\L_\r^\m \L_\s^\n \pd^{[\r} a^{\s]}=\L_\r^\m \L_\s^\n f^{\r\s}$ with $\pdb^\m=\L_\r^\m \pd^\r$ and $\ab^\n=\L_\s^\n a^\s$. Using $a^\m=\Lb_\n^\m \ab^\n$ and $\pd^\m=\Lb_\n^\m \pdb^\n$, we can decompose the $O(\pd a)$ electromagnetic field into parallel and perpendicular parts in the LRF. To proceed, we define $\L_\n^\m=\(\L_u+\L_\pp+\L_b\)_\n^\m,\,\Lb_\n^\m=\(\Lb_u+\Lb_\pp+\Lb_b\)_\n^\m$ with $\L_{\n,u}^\m\equiv\(u^\n,\,-0_k^\n,\,-0^\n\)^T,\,\Lb_{\n,u}^\m\equiv\(u^\m,\,0_k^\m,\,0^\m\)$, $\L_{\n,\pp}^\m\equiv\(0^\n,\,-w_k^\n,\,-0^\n\)^T,\,\Lb_{\n,\pp}^\m\equiv\(0^\m,\,w_k^\m,\,0^\m\)$ and $\L_{\n,b}^\m\equiv\(0^\n,\,-0_k^\n,\,-b^\n\)^T,\,\Lb_{\n,b}^\m\equiv\(0^\m,\,0_k^\m,\,b^\m\)$. We then have
\begin{align}
  f^{\m\n}=\pd^{[\m} a^{\n]}=\Lb_\r^\m \Lb_\s^\n \pdb^{[\r}\ab^{\s]}=&\Big[\(\Lb_{\r,u}^\m \Lb_{\s,b}^\n+\Lb_{\r,b}^\m \Lb_{\s,u}^\n\)+\(\Lb_{\r,u}^\m \Lb_{\s,\pp}^\n+\Lb_{\r,\pp}^\m \Lb_{\s,u}^\n\)\no
  &+\Lb_{\r,\pp}^\m \Lb_{\s,\pp}^\n+\(\Lb_{\r,b}^\m \Lb_{\s,\pp}^\n+\Lb_{\r,\pp}^\m \Lb_{\s,b}^\n\)\Big]\pdb^{[\r}\ab^{\s]}\no
  \equiv& f_{E\pr}^{\m\n} + f_{E\pp}^{\m\n} + f_{\cB\pr}^{\m\n} + f_{\cB\pp}^{\m\n}\,,
\end{align}
where the four cases in the last line are defined by the four terms in the bracket of the second to the last equality. Simialrly,
\begin{align}
  f_{\m\n}=\pd_{[\m} a_{\n]}=\L_\m^\r \L_\n^\s \pdb_{[\r}\ab_{\s]}=&\Big[\(\L_{\m,u}^\r \L_{\n,b}^\s+\L_{\m,b}^\r \L_{\n,u}^\s\)+\(\L_{\m,u}^\r \L_{\n,\pp}^\s+\L_{\m,\pp}^\r \L_{\n,u}^\s\)\no
  &+\L_{\m,\pp}^\r \L_{\n,\pp}^\s+\(\L_{\m,b}^\r \L_{\n,\pp}^\s+\L_{\m,\pp}^\r \L_{\n,b}^\s\)\Big]\pdb_{[\r}\ab_{\s]}\no
  \equiv& f^{E\pr}_{\m\n} + f^{E\pp}_{\m\n} + f^{\cB\pr}_{\m\n} + f^{\cB\pp}_{\m\n}\,.
\end{align}
Thus, we get $E_{\pr,\pp}^\m=f_{E\pr,\pp}^{\m\n} u_\n,\;\cB_{\pr,\pp}^\m=\frac{1}{2}\e^{\m\n\r\s}u_\n f_{\r\s}^{\cB\pr,\pp}$.

Accordingly, in momentum space, one has
\begin{align}\label{f_ms}
  f_{\m\n}^{E\pr}\to& i\(\qb_0\ab_3-\qb_3\ab_0\)u_{[\m} b_{\n]}\,,\qquad
  f^{\m\n}_{E\pr}\to i\(\qb_0\ab_3-\qb_3\ab_0\)u^{[\m} b^{\n]}\,,\no
  f_{\m\n}^{E\pp}\to& i\(\qb_0\ab_k-\qb_k\ab_0\)u_{[\m} w_{\n]}^k\,,\qquad
  f^{\m\n}_{E\pp}\to i\(\qb_0\ab_k-\qb_k\ab_0\)u^{[\m} w^{\n]}_k\,,\no
  f_{\m\n}^{\cB\pr}\to& i\(\qb_1\ab_2-\qb_2\ab_1\)w_{[\m}^1 w_{\n]}^2\,,\qquad
  f^{\m\n}_{\cB\pr}\to i\(\qb_1\ab_2-\qb_2\ab_1\)w^{[\m}_1 w^{\n]}_2\,,\no
  f_{\m\n}^{\cB\pp}\to& i\(\qb_3\ab_k-\qb_k\ab_3\)b_{[\m} w_{\n]}^k\,,\qquad
  f^{\m\n}_{\cB\pp}\to i\(\qb_3\ab_k-\qb_k\ab_3\)b^{[\m} w^{\n]}_k\,.
\end{align}

Note we've suppressed helicity $s$ in the notation of the perturbative gauge field $a_s^\l$ for simplicity. Upon summation $a=(a_++a_-)/2$ and subtraction $a_5=(a_+-a_-)/2$, one can easily generalize the above results into vectorial and axial gauge fields.

\section{Solving covariant CKE}\label{app_solution}

For the constraints in $E_\pr$ case, upon Fourier transformation, we have $f_{\m\n}^{E\pr}\to i\(\qb_0\ab_3-\qb_3\ab_0\)u_{[\m} b_{\n]}$. Also, $\pd_\m^{E\pr} j_a\to i\hat q_\m \(\ab_0\frac{\pd j_a}{\pd \ab_0}+\ab_3\frac{\pd j_a}{\pd \ab_3}\)$ using \eqref{sp_pd}. Then one gets constraints on $j_a$ from the left hand side(LHS) and right hand side(RHS) of \eqref{src1},
\begin{align}
  \text{LHS}=&q\cdot(u+b)\(\ab_0\frac{\pd j_a}{\pd \ab_0}+\ab_3\frac{\pd j_a}{\pd \ab_3}\)\no
  \text{RHS}=&\(\qb_0\ab_3-\qb_3\ab_0\)u_{[\m} b_{\n]}(u+b)^\m \frac{\pd}{\pd p_\n}j=\(\qb_0\ab_3-\qb_3\ab_0\)\(\frac{\pd j}{\pd\(p\cdot u\)}-\frac{\pd j}{\pd\(p\cdot b\)}\)\,.
\end{align}
In \eqref{src2}, using $\e^{\m\n\r\s} \qh_\r (u+b)_\s=\e^{\m\n\r\s} q\cdot(u+b)u_\r b_\s$, one gets
\begin{align}
  \text{LHS}=&\e^{\m\n\r\s} u_\r b_\s q\cdot(u+b)\(\ab_0\frac{\pd j_a}{\pd \ab_0}+\ab_3\frac{\pd j_a}{\pd \ab_3}\)\no
  \text{RHS}=&\e^{\m\n\r\s}\(\qb_0\ab_3-\qb_3\ab_0\)u_{[\r} b_{\l]}(u+b)_\s \frac{\pd}{\pd p_\l}j\no
  =&\e^{\m\n\r\s} u_\r b_\s \(\qb_0\ab_3-\qb_3\ab_0\) \(\frac{\pd j}{\pd\(p\cdot u\)}-\frac{\pd j}{\pd\(p\cdot b\)}\).
\end{align}
We can see both \eqref{src1} and \eqref{src2} give \eqref{cj_epr}
as the constraint on $j_a$ in $E_\pr$ case. 

For $E_\pp$ and $\cB_\pp$, noting
\begin{align}
  \pd_\m^{E\pp} j_a\;\to\; i\(u_\m\qb_0 \ab_k\frac{\pd j_a}{\pd \ab_k}+q_\m^T \ab_0 \frac{\pd j_a}{\pd \ab_0}\)=i\qb_0\(u_\m \ab_k\frac{\pd j_a}{\pd \ab_k}+a_\m^T \frac{\pd j_a}{\pd \ab_0}\)+f_{\m\n}^{E\pp} u^\n \frac{\pd j_a}{\pd \ab_0}\,,\no
  \pd_\m^{\cB\pp} j_a\;\to\; i\(b_\m\qb_3 \ab_k\frac{\pd j_a}{\pd \ab_k}+q_\m^T \ab_3 \frac{\pd j_a}{\pd \ab_3}\)=i\qb_3\(b_\m \ab_k\frac{\pd j_a}{\pd \ab_k}+a_\m^T \frac{\pd j_a}{\pd \ab_3}\)-f_{\m\n}^{\cB\pp} b^\n \frac{\pd j_a}{\pd \ab_3}\,.
\end{align}
The scalar equations \eqref{eom1E} and \eqref{eom2E} give
\begin{align}
  \eqref{eom1E}\quad\xrightarrow{E\pp}\quad& p\cdot u_\pp\, p\cdot (u+b)\(\frac{\pd }{\pd(p\cdot u)}G_1+G_2\)+p\cdot u_\pp  G_3=0\label{eom1_drift}\\
  \xrightarrow{\cB\pp}\quad& \frac{p\cdot \cB_\pp}{B} p\cdot (u+b)\(\frac{\pd }{\pd(p\cdot u)}H_1+H_2\)+\frac{p\cdot \cB_\pp}{B}  H_3=0\label{eom1_tilt}
\end{align}
\begin{align}
  \eqref{eom2E}\quad\xrightarrow{E\pp}\quad& \pd_\m^{E\pp} j_{(0) a}^{\m }-F_{\m\l }\frac{\pd}{\pd p_\l}j_{E\pp }^{\m }-f_{\m\n}^{E\pp} \frac{\pd }{\pd p_{\n }}j_{(0)}^{\m }\no
  &=i\qb_0\ab_k\frac{\pd j_a}{\pd \ab_k}+B\e_{\m \l \a \b }b^{\a }u^{\b }\frac{\e^{\m \n \r \s } f_{\n\r}^{E\pp}  b_{\s }}{2B}\frac{2p_T^{\l }}{B}G_3-(u+b)^{\m } f_{\m\n}^{E\pp}\frac{2p_T^{\n }}{B}j\no
  &=i\qb_0\ab_k\frac{\pd j_a}{\pd \ab_k}+p_T^{[\r } u^{\n ]}f_{\n\r}^{E\pp}  \frac{1}{B}G_3-p_T^{[\n } u^{\m ]}f_{\m\n}^{E\pp} \frac{1}{B}j=0\,,\label{eom2_drift}
\end{align}
\begin{align}
  \xrightarrow{\cB\pp}\quad& \pd_\m^{\cB\pp} j_{(0) a}^{\m }-F_{\m\l }\frac{\pd}{\pd p_\l}j_{\cB\pp }^{\m }-f_{\m\n}^{\cB\pp} \frac{\pd }{\pd p_{\n }}j_{(0)}^{\m }\no
  &=-i\qb_3\ab_k\frac{\pd j_a}{\pd \ab_k}+B\e_{\m \l \a \b }b^{\a }u^{\b }\frac{\e^{\m \n \r \s } f_{\n\r}^{\cB\pp}  u_{\s }}{2B}\frac{2p_T^{\l }}{B}H_3-(u+b)^{\m } f_{\m\n}^{\cB\pp}\frac{2p_T^{\n }}{B}j\no
  &=-i\qb_3\ab_k\frac{\pd j_a}{\pd \ab_k}+p_T^{[\r } b^{\n ]}f_{\n\r}^{\cB\pp}  \frac{1}{B}H_3-p_T^{[\n } b^{\m ]}f_{\m\n}^{\cB\pp} \frac{1}{B}j=0\,,\label{eom2_tilt}
\end{align}
where we have used $b^\m f_{\m\n}^{E\pp}=u^\m f_{\m\n}^{\cB\pp}=0$ in \eqref{eom2_drift} and \eqref{eom2_tilt}. 
One finds \eqref{eom1_drift} to \eqref{eom2_tilt} are satisfied by
\begin{align}
  \ab_k\frac{\pd j_a}{\pd \ab_k}=0\,,\qquad  &G_1=G_3=j\,,\qquad G_2\propto \d(p\cdot(u+b))\,,\no
  &H_1=H_3=j\,,\qquad H_2\propto \d(p\cdot(u+b))\,.
\end{align}

The anti-symmetric tensor equation \eqref{eom3E} is simplified as follows. The two parts on the LHS give
\begin{align}\label{lhs_E}
  &p^{[\m }j_{E\pp }^{\n ]}=p\cdot u_\pp p^{[\m }(u+b)^{\n ]}\(\frac{\pd }{\pd(p\cdot u)}G_1+G_2\)+p^{[\m } u_\pp ^{\n ]}G_3\no
  &=p\cdot u_\pp \(p_T^{[\m }(u+b)^{\n ]}-p\cdot(u+b)b^{[\m}u^{\n]}\)\(\frac{\pd }{\pd(p\cdot u)}G_1+G_2\)+p_T^{[\m } u_\pp ^{\n ]}G_3+p\cdot u(u+b)^{[\m } u_\pp ^{\n ]}G_3\,,
\end{align}
\begin{align}\label{lhs_B}
  &p^{[\m }j_{\cB\pp }^{\n ]}=\frac{p\cdot \cB_\pp}{B} p^{[\m }(u+b)^{\n ]}\(\frac{\pd }{\pd(p\cdot u)}H_1+H_2\)+\frac{1}{B}p^{[\m } \cB_\pp^{\n ]}H_3\no
  &=\frac{p\cdot \cB_\pp}{B} \(p_T^{[\m }(u+b)^{\n ]}-p\cdot(u+b)b^{[\m}u^{\n]}\)\(\frac{\pd }{\pd(p\cdot u)}H_1+H_2\)+\frac{1}{B}p_T^{[\m } \cB_\pp^{\n ]}H_3+\frac{1}{B}p\cdot u(u+b)^{[\m } \cB_\pp^{\n ]}H_3\,,
\end{align}
\begin{align}\label{lhs_E1}
  &\frac{1}{2}\e^{\m \n \r \s }F_{\r\l}\frac{\pd}{\pd p_\l} j^{E\pp }_{\s }=-\frac{1}{2}\e^{\m \n \r \s }B\e_{\r \l \a \b }b^{\a }u^{\b }\frac{\pd }{\pd p_{\l }}\[p\cdot u_\pp (u+b)_{\s }\(\frac{\pd }{\pd(p\cdot u)}G_1+G_2\)+u_\s^\pp  G_3\]\no
  &=\(p\cdot u_\pp p_T^{[\m }(u+b)^{\n ]}+\frac{B}{2}u_\pp ^{[\m }(u+b)^{\n ]}\)\(\frac{\pd }{\pd(p\cdot u)}G_1+G_2\)+b^{[\m }u^{\n ]}p\cdot u_\pp G_3\,,
\end{align}
\begin{align}\label{lhs_B1}
  &\frac{1}{2}\e^{\m \n \r \s }F_{\r\l}\frac{\pd}{\pd p_\l} j^{\cB\pp }_{\s }=-\frac{1}{2}\e^{\m \n \r \s }B\e_{\r \l \a \b }b^{\a }u^{\b }\frac{\pd }{\pd p_{\l }}\[\frac{p\cdot \cB_\pp}{B} (u+b)_{\s }\(\frac{\pd }{\pd(p\cdot u)}H_1+H_2\)+\frac{\cB_\s^\pp}{B}H_3\]\no
  &=\(\frac{p\cdot \cB_\pp}{B} p_T^{[\m }(u+b)^{\n ]}+\frac{1}{2}\cB_\pp ^{[\m }(u+b)^{\n ]}\)\(\frac{\pd }{\pd(p\cdot u)}H_1+H_2\)+b^{[\m }u^{\n ]}\frac{p\cdot \cB_\pp}{B} H_3\,.
\end{align}
Noting $f_{\r\l}^{E\pp} b^\l=f_{\r\l}^{\cB\pp} u^\l=0$, the inhomogeneous terms on the RHS write
\begin{align}\label{rhs_E}
  &-\frac{1}{2}\e^{\m \n \r \s }\(\pd_\r^{E\pp} j_{\s }^{a}-f_{\r\l}^{E\pp} \frac{\pd }{\pd p_{\l }}j_{\s }^{(0)}\)\no
  &=-\frac{1}{2}\e^{\m \n \r \s }\[i\qb_0\(u_\r\ab_k \frac{\pd j_a}{\pd \ab_k}+ a_\r^T\frac{\pd j_a}{\pd \ab_0}\)+f_{\r\l}^{E\pp} u^{\l }\frac{\pd j_a}{\pd \ab_0}\](u+b)_\s+\frac{1}{2}\e^{\m \n \r \s }f_{\r\l}^{E\pp}(u+b)_{\s }\(u^{\l }\frac{\pd }{\pd(p\cdot u)}+\frac{2p_T^{\l }}{B}\)j\no
  &=-\frac{i}{2}\e^{\m \n \r \s }\( u_\r b_\s \frac{\pd j_a}{\pd \ab_k}+ w_\r^k(u+b)_\s\frac{\pd j_a}{\pd \ab_0}\)\qb_0 \ab_k+\frac{B}{2}(u+b)^{[\m }u_\pp ^{\nu ]}\(\frac{\pd j}{\pd(p\cdot u)}-\frac{\pd j_a}{\pd \ab_0}\)+p_T^{[\m} u_\pp ^{\n]}j\,,
\end{align}
\begin{align}\label{rhs_B}
  &-\frac{1}{2}\e^{\m \n \r \s }\(\pd_\r^{\cB\pp} j_{\s }^{a}-f_{\r\l}^{\cB\pp} \frac{\pd }{\pd p_{\l }}j_{\s }^{(0)}\)\no
  &=-\frac{1}{2}\e^{\m \n \r \s }\[i\qb_3\(b_\r\ab_k \frac{\pd j_a}{\pd \ab_k}+a_\r^T\frac{\pd j_a}{\pd \ab_3}\)-f_{\r\l}^{\cB\pp} b^{\l }\frac{\pd j_a}{\pd \ab_3}\](u+b)_\s+\frac{1}{2}\e^{\m \n \r \s }f_{\r\l}^{\cB\pp}(u+b)_{\s }\(b^{\l }\frac{\pd }{\pd(p\cdot b)}+\frac{2p_T^{\l }}{B}\)j\no
  &=-\frac{i}{2}\e^{\m \n \r \s }\(b_\r u_\s \frac{\pd j_a}{\pd \ab_k}+ w_\r^k(u+b)_\s\frac{\pd j_a}{\pd \ab_3}\)\qb_3 \ab_k+\frac{1}{2}(u+b)^{[\m }\cB_\pp^{\nu ]}\(\frac{\pd j}{\pd(p\cdot b)}+\frac{\pd j_a}{\pd \ab_3}\)+\frac{1}{B}p_T^{[\m} \cB_\pp^{\n]}j\,,
\end{align}
where we have used the following identities,
\begin{align}\label{identity}
  \e^{\m \n \r \s } f_{\r\l}^{E\pp}  (u+b)_{\s } u^{\l }=&B(u+b)^{[\m}u_\pp ^{\n]}\,,\qquad\,
  \e^{\m \n \r \s } f_{\r\l}^{E\pp}  (u+b)_{\s } p_T^{\l }=B p_T^{[\m }u_\pp ^{\n]}\,,\no
  \e^{\m \n \r \s } f_{\r\l}^{\cB\pp}  (u+b)_{\s } b^{\l }=&(u+b)^{[\m}\cB_\pp^{\n]}\,,\qquad\quad
  \e^{\m \n \r \s } f_{\r\l}^{\cB\pp}  (u+b)_{\s } p_T^{\l }=p_T^{[\m }\cB_\pp^{\n]}\,.
\end{align}
We collect the terms from \eqref{lhs_E} to \eqref{rhs_B} and group them into $\e^{\m \n \r \s }u_\r b_\s$, $\e^{\m \n \r \s }w_\r^k(u+b)_\s$, $b^{[\m}u^{\n]}$, $p_T^{[\m}(u+b)^{\n]}$, $(u+b)^{[\m}u_\pp ^{\n]}$, $p_T^{[\m}u_\pp ^{\n]}$, $(u+b)^{[\m}\cB_\pp ^{\n]}$ and $p_T^{[\m}\cB_\pp ^{\n]}$ terms to fix $G_n,H_n$ by comparing the coefficients of the groups. For $\e^{\m \n \r \s }u_\r b_\s$ and $\e^{\m \n \r \s }w_\r^k(u+b)_\s$ terms, one gets
\begin{align}
  \frac{\pd j_a}{\pd \ab_k}\ab_k\(\qb_0-\qb_3\)=0\,,\qquad \(\frac{\pd j_a}{\pd \ab_0}\qb_0+\frac{\pd j_a}{\pd \ab_3}\qb_3\)\ab_k=0\,.
\end{align}
For $b^{[\m}u^{\n]}$ terms, one gets
\begin{align}
  -p\cdot u_\pp p\cdot (u+b)\(\frac{\pd }{\pd(p\cdot u)}G_1+G_2\)=p\cdot u_\pp G_3\,,\no
  -p\cdot \cB_\pp p\cdot (u+b)\(\frac{\pd }{\pd(p\cdot u)}H_1+H_2\)=p\cdot \cB_\pp H_3\,,
\end{align}
which holds by $G_1=G_3=H_1=H_3=j$ and $G_2,H_2\propto \d(p\cdot(u+b))$. The coefficients of $p_T^{[\m}(u+b)^{\n]}$ in two sides cancel out automatically. 
For the $(u+b)^{[\m}u_\pp ^{\n ]}$ and $(u+b)^{[\m}\cB_\pp ^{\n ]}$ terms, we get
\begin{align}
  p\cdot u\,G_3=-\frac{B}{2}\(\frac{\pd }{\pd(p\cdot u)}G_1+G_2\)-\frac{B}{2}\frac{\pd j_a}{\pd \ab_0}+\frac{B}{2}\frac{\pd }{\pd(p\cdot u)}j\,,\no
  p\cdot u\,H_3=-\frac{B}{2}\(\frac{\pd }{\pd(p\cdot u)}H_1+H_2\)+\frac{B}{2}\frac{\pd j_a}{\pd \ab_3}+\frac{B}{2}\frac{\pd }{\pd(p\cdot b)}j\,.
\end{align}
which, by $G_1=G_3=H_1=H_3=j$, gives
\begin{align}
  G_2=-\frac{2p\cdot u}{B}j-\frac{\pd j_a}{\pd \ab_0}\,,\qquad H_2=\frac{\pd j}{\pd(p\cdot b)}-\frac{\pd j}{\pd(p\cdot u)}-\frac{2p\cdot u}{B}j+\frac{\pd j_a}{\pd \ab_3}\,.
\end{align}
The coefficients of $p_T^{[\m } u_\pp ^{\n ]}$ and $p_T^{[\m } \cB_\pp ^{\n ]}$ give $G_3=H_3=j$. 

\section{Functional differentiation}\label{app_functional}

In momentum space, the functional differential operators write
\begin{align}
  \frac{\d a_\m(k)}{\d a_\n(q)}=\frac{\d a_\m^5(k)}{\d a_\n^5(q)}=\int d^4k\,\d_\m^\n\, \d^{(4)}(k-q)\,,\qquad \frac{\d a_\m(k)}{\d a_\n^5(q)}=\frac{\d a_\m^5(k)}{\d a_\n(q)}=0\,.
\end{align}
The momentum integration involving $\d$ function is easy, which gives
\begin{align}
  \cG_{VV}^{\m\n}(q)=\frac{\d\cJ^\m(k)}{\d a_\n(q)}=\frac{\d^2 \cL}{\d a_\n(q)\, \d a_\m(-k)}=\frac{\d^2 \cL}{\d a_\m(-k)\, \d a_\n(q)}=\frac{\d\cJ^\n(-q)}{\d a_\m(-k)}=\cG_{VV}^{\n\m}(-k)\,,
\end{align}
where we have taken $\cG_{VV}$  for example. Note $q$ as the variable of the function $\cG_{VV}^{\m\n}(q)$ can be renamed $k$, one then gets the derivative symmetry $\cG_{VV}^{\m\n}(k)=\cG_{VV}^{\n\m}(-k)$. In the following processes, we will omit the trivial momentum integration to be concise. To get \eqref{grn_0}, the variation of $J_{(0)}^{\m}$ and $J_{(0)5}^{\m}$ with respect to $a_\l$ and $a_\l^5$ involves
\begin{align}
  \frac{\d\D\m}{\d a_\l}=&\frac{\d\D\m_5}{\d a_\l^5}=\frac{\(u^\l\qb_3 + b^\l\qb_0\)\qb_3}{\qb_0^2-\qb_3^2}\,,\no
  \frac{\d\D\m_5}{\d a_\l}=&\frac{\d\D\m}{\d a_\l^5}=\frac{\(u^\l\qb_3 + b^\l\qb_0\)\qb_0}{\qb_0^2-\qb_3^2}\,.
\end{align}

To derive \eqref{grn_1}, we use $U^\m=\frac{1}{2}\e^{\m \n \r \s }b_\n f_{\r\s}^{E\pp}$, $U_5^\m=\frac{1}{2}\e^{\m \n \r \s }b_\n f_{\r\s}^{E\pp 5}$, $\cB_{\pr/\pp}^\m=\frac{1}{2}\e^{\m\n\r\s}u_\n f_{\r\s}^{\cB\pr/\pp}$ and $\cB_{\pr/\pp 5}^\m=\frac{1}{2}\e^{\m\n\r\s}u_\n f_{\r\s}^{\cB\pr/\pp 5}$ together with \eqref{f_ms}. The variation of $J_{\cA }^{\m}$ and $J_{\cA 5}^{\m}$ with respect to $a_\l$ and $a_\l^5$ involves
\begin{align}
  \frac{\d U^\m}{\d a_\l}&=\frac{1}{2}\e^{\m \n \r \s }b_\n i\(\qb_0\frac{\d \ab_k}{\d a_\l}-\qb_k\frac{\d \ab_0}{\d a_\l}\)u_{[\r} w_{\s]}^k\no
  &=i\e^{\m \n \r \s }\(\qb_0 w_k^\l+\qb_k u^\l\)w_\n^k u_\r b_\s
  =i\(-\e^{\m \l \r \s }\qb_0 + \e^{\m \n \r \s } u^\l q_\n^T\) u_\r b_\s, \no
  \frac{\d\(b\cdot \cB_\pr\)}{\d a_\l}&=\frac{1}{2}\e^{\s \r \m \n }b_\s u_\r i\(\qb_1\frac{\d \ab_2}{\d a_\l}-\qb_2\frac{\d \ab_1}{\d a_\l}\)w_{[\m}^1 w_{\n]}^2\no
  &=i\e^{\m \n \r \s }\(\qb_1 w_2^\l-\qb_2 w_1^\l\)w_\m^1 w_\n^2 u_\r b_\s
  =i\e^{\l \n \r \s } q_\n^T u_\r b_\s\,,\no
  \frac{\d \cB_\pp^\m}{\d a_\l}&=\frac{1}{2}\e^{\m \n \r \s }u_\n i\(\qb_3\frac{\d \ab_k}{\d a_\l}-\qb_k\frac{\d \ab_3}{\d a_\l}\)b_{[\r} w_{\s]}^k\no
  &=i\e^{\m \n \r \s }\(-\qb_3 w_k^\l+\qb_k b^\l\)w_\n^k u_\r b_\s
  =i\(\e^{\m \l \r \s }\qb_3 + \e^{\m \n \r \s } b^\l q_\n^T\) u_\r b_\s\,,\no
  \frac{\d\cB_\pr^\m}{\d a_\l}&=\frac{\d\(-b\cdot\cB_\pr b^\m\)}{\d a_\l}\to -i\e^{\l \n \r \s } q_\n^T u_\r b_\s b^\m\,,\no
  \frac{\d U_5^\m}{\d a_\l^5}=\frac{\d U^\m}{\d a_\l}\,,&\qquad
  \frac{\d\(b\cdot \cB_{\pr 5}\)}{\d a_\l^5}=\frac{\d\(b\cdot \cB_\pr\)}{\d a_\l}\,,\qquad
  \frac{\d \cB_{\pp 5}^\m}{\d a_\l^5}=\frac{\d \cB_\pp^\m}{\d a_\l}\,,\qquad
  \frac{\d\cB_{\pr 5}^\m}{\d a_\l^5}=\frac{\d\cB_\pr^\m}{\d a_\l}\,,\no
  \frac{\d U_5^\m}{\d a_\l}=\frac{\d U^\m}{\d a_\l^5}=&\frac{\d\(b\cdot \cB_{\pr 5}\)}{\d a_\l}=\frac{\d\(b\cdot \cB_\pr\)}{\d a_\l^5}=\frac{\d \cB_{\pp 5}^\m}{\d a_\l}=\frac{\d \cB_\pp^\m}{\d a_\l^5}=\frac{\d\(\cB_{\pr 5}^\m\)}{\d a_\l}=\frac{\d\(\cB_\pr^\m\)}{\d a_\l^5}=0\,,
\end{align}
where we have used $\e^{\m \n \r \s } w_k^\l w_\n^k u_\r b_\s=-\e^{\m \l \r \s } u_\r b_\s$ and $\e^{\m \n \r \s }w_i^\l w_\m^k w_\n^i u_\r b_\s=\e^{\l \n \r \s } w_\n^k u_\r b_\s$.

\bibliographystyle{unsrt}
\bibliography{correlator_CKT.bib}
  
\end{document}